\documentclass[prb,twocolumn,showpacs,preprintnumbers,amsmath,aps]{revtex4}
\usepackage{graphicx,hyperref}
\usepackage{bm}
\usepackage{subfigure}

\newcommand\vect[1]{ \boldsymbol{ #1}}
\def\nbC{{\mathchoice {\setbox0=\hbox{$\displaystyle\rm C$}%
\hbox{\hbox to0pt{\kern0.4\wd0\vrule height0.9\ht0\hss}\box0}}
{\setbox0=\hbox{$\textstyle\rm C$}\hbox{\hbox
to0pt{\kern0.4\wd0\vrule height0.9\ht0\hss}\box0}}
{\setbox0=\hbox{$\scriptstyle\rm C$}\hbox{\hbox
to0pt{\kern0.4\wd0\vrule height0.9\ht0\hss}\box0}}
{\setbox0=\hbox{$\scriptscriptstyle\rm C$}\hbox{\hbox
to0pt{\kern0.4\wd0\vrule height0.9\ht0\hss}\box0}}}}
\def\nbQ{{\mathchoice {\setbox0=\hbox{$\displaystyle\rm
Q$}\hbox{\raise
0.15\ht0\hbox to0pt{\kern0.4\wd0\vrule height0.8\ht0\hss}\box0}}
{\setbox0=\hbox{$\textstyle\rm Q$}\hbox{\raise
0.15\ht0\hbox to0pt{\kern0.4\wd0\vrule height0.8\ht0\hss}\box0}}
{\setbox0=\hbox{$\scriptstyle\rm Q$}\hbox{\raise
0.15\ht0\hbox to0pt{\kern0.4\wd0\vrule height0.7\ht0\hss}\box0}}
{\setbox0=\hbox{$\scriptscriptstyle\rm Q$}\hbox{\raise
0.15\ht0\hbox to0pt{\kern0.4\wd0\vrule height0.7\ht0\hss}\box0}}}}
\def\nbT{{\mathchoice {\setbox0=\hbox{$\displaystyle\rm
T$}\hbox{\hbox to0pt{\kern0.3\wd0\vrule height0.9\ht0\hss}\box0}}
{\setbox0=\hbox{$\textstyle\rm T$}\hbox{\hbox
to0pt{\kern0.3\wd0\vrule height0.9\ht0\hss}\box0}}
{\setbox0=\hbox{$\scriptstyle\rm T$}\hbox{\hbox
to0pt{\kern0.3\wd0\vrule height0.9\ht0\hss}\box0}}
{\setbox0=\hbox{$\scriptscriptstyle\rm T$}\hbox{\hbox
to0pt{\kern0.3\wd0\vrule height0.9\ht0\hss}\box0}}}}
\def\nbS{{\mathchoice
{\setbox0=\hbox{$\displaystyle     \rm S$}\hbox{\raise0.5\ht0%
\hbox to0pt{\kern0.35\wd0\vrule height0.45\ht0\hss}\hbox
to0pt{\kern0.55\wd0\vrule height0.5\ht0\hss}\box0}}
{\setbox0=\hbox{$\textstyle        \rm S$}\hbox{\raise0.5\ht0%
\hbox to0pt{\kern0.35\wd0\vrule height0.45\ht0\hss}\hbox
to0pt{\kern0.55\wd0\vrule height0.5\ht0\hss}\box0}}
{\setbox0=\hbox{$\scriptstyle      \rm S$}\hbox{\raise0.5\ht0%
\hboxto0pt{\kern0.35\wd0\vrule height0.45\ht0\hss}\raise0.05\ht0%
\hbox to0pt{\kern0.5\wd0\vrule height0.45\ht0\hss}\box0}}
{\setbox0=\hbox{$\scriptscriptstyle\rm S$}\hbox{\raise0.5\ht0%
\hboxto0pt{\kern0.4\wd0\vrule height0.45\ht0\hss}\raise0.05\ht0%
\hbox to0pt{\kern0.55\wd0\vrule height0.45\ht0\hss}\box0}}}}
\def\nbZ{{\mathchoice {\hbox{$\sf\textstyle Z\kern-0.4em Z$}}
{\hbox{$\sf\textstyle Z\kern-0.4em Z$}}
{\hbox{$\sf\scriptstyle Z\kern-0.3em Z$}}
{\hbox{$\sf\scriptscriptstyle Z\kern-0.2em Z$}}}}

\begin{document}

\title{Hierarchical Reference Theory of critical fluids in disordered porous media}

\author{Gilles Tarjus} \email{tarjus@lptl.jussieu.fr}
\affiliation{LPTMC, CNRS-UMR 7600, Universit\'e Pierre et Marie Curie,
bo\^ite 121, 4 Place Jussieu, 75252 Paris c\'edex 05, France}

\author{Martin-Luc Rosinberg} \email{mlr@lptl.jussieu.fr}
\affiliation{LPTMC, CNRS-UMR 7600, Universit\'e Pierre et Marie Curie,
bo\^ite 121, 4 Place Jussieu, 75252 Paris c\'edex 05, France}

\author{Edouard Kierlik} \email{kierk@lptl.jussieu.fr}
\affiliation{LPTMC, CNRS-UMR 7600, Universit\'e Pierre et Marie Curie,
bo\^ite 121, 4 Place Jussieu, 75252 Paris c\'edex 05, France}

\author{Matthieu Tissier} \email{tissier@lptmc.jussieu.fr}
\affiliation{LPTMC, CNRS-UMR 7600, Universit\'e Pierre et Marie Curie,
bo\^ite 121, 4 Place Jussieu, 75252 Paris c\'edex 05, France}

\date{\today}

\begin{abstract}
We consider the equilibrium behavior of fluids imbibed in disordered mesoporous media, including their gas-liquid critical point when present. Our starting points are on the one hand a description of the fluid/solid-matrix system as a quenched-annealed mixture and on the other hand the Hierarchical Reference Theory (HRT) developed by A. Parola and L. Reatto to cope with density fluctuations on all length scales. The formalism combines liquid-state statistical mechanics and the theory of systems in the presence of quenched disorder. A straightforward implementation of the HRT to the quenched-annealed mixture is shown to lead to unsatisfactory results, while indicating that the critical behavior of the system is in the same universality class as that of the random-field Ising model. After a detour via the field-theoretical renormalization group approach of the latter model, we finally lay out the foundations for a proper HRT of fluids in a disordered porous material.
\end{abstract}


\maketitle

\section{Introduction}
\label{sec:introduction}

The  HRT developed by A. Parola and L. Reatto\cite{parola-reatto85,HRTreview,HRTsmooth} combines liquid-state statistical mechanics and nonperturbative renormalization group (RG) formalism. As such, it has proven a powerful tool for describing the equilibrium structure and the phase behavior of simple fluids and fluid mixtures in a way that properly incorporates the long wavelength fluctuations present in the vicinity of the critical point(s)\cite{parola-reatto85,HRTreview,HRTsmooth,HRTapplic,HRTkahl02,HRTverso06}.

The purpose of this article is to show how the HRT can be generalized to deal with fluids in disordered mesoporous media. The critical behavior of such systems is indeed an unsettled question. It was suggested by de Gennes and Brochard\cite{degennes} that fluids imbibed by very dilute disordered porous media like silica gels represent an experimental realization of the random-field Ising model (RFIM). The rationale is as follows: as the solid attracts preferentially one phase of the fluid (usually the liquid for a single-component fluid and one of the species for a mixture), there is a shift of density or concentration away from the solid; this can be described as the result of a perturbation to the chemical potential which, due to the disordered structure of the gel, is random in space. At a coarse-grained level, the fluid near criticality in the presence of a gel can then be modeled by a lattice gas with a random chemical potential which, provided the correlation length of the fluid is much larger than that characterizing the structure of the gel, can be taken as uncorrelated from site to site. Going to the Ising-spin representation of the lattice gas thus leads to a ferromagnetic Ising model in a random magnetic field. The random-field Ising model (RFIM)\cite{nattermann98} is an archetype of statistical-mechanical model in the presence of ``quenched disorder''. The quenched disorder in the fluid case is associated with the structure of the solid matrix of the porous medium.

It has however proven extremely difficult to test the de Gennes-Brochard suggestion in experiments and in computer simulations.  This is due to strong crossover effects to bulk-like behavior when the effect of the quenched disorder is very weak\cite{exp-chanbeta} and to extremely slow dynamics in general\cite{exp-knobler,exp-wiltzius,exp-canell,exp-chan,exp-puech}. Actually the vast majority of experimental studies appear to be out of equilibrium and characterized by the presence of hysteresis effects. Such an hysteresis is routinely found in adsorption experiments\cite{adsorption-review} where the filling of the porous medium by the fluid (the ``capillary condensation'') and the draining of the medium (the ``capillary evaporation'') do not take place at the same chemical potential. Phase transition could still be observed in very open porous materials such as light aerogels\cite{exp-wolf}, but these are out-of-equilibrium phase transitions\cite{kierk-hysteresis,detche-hysteresis}. 

The lack of strong experimental evidence showing the equivalence between the critical behavior of fluids in disordered mesoporous media and that of the RFIM could come from several features that are missing in the simple RFIM: the fact that the porous material imposes a correlated disorder on the fluid or that the distribution of random chemical potentials is strongly asymmetric\cite{maritan91}, the effect of confinement and/or of dilution, etc. On the other hand, a recent series of Monte Carlo studies of several models of fluids in disordered media\cite{binder} has concluded in favor of the de Gennes-Brochard conjecture.

The HRT appears as a promising candidate framework to tackle the behavior of fluids in disordered mesoporous materials. It is indeed able to provide a nonperturbative implementation of the RG in the appropriate regime near the critical point. It also allows one to start with, and maintain, a microscopic description of the system, including in the present case confinement and/or dilution, correlated asymmetric disorder, and wetting phenomena.  One should finally add that the critical behavior of the RFIM has itself been under debate since its introduction in the mid-seventies\cite{imry-ma75} and that the standard perturbative RG approaches have been shown to be inadequate to describe the critical behavior of the system\cite{nattermann98}. Nonperturbative RG methods then offer a powerful alternative\cite{tarjus04,tissier06,tissier11}.

The outline of this paper is as follows. In section~\ref{sec:HRTbrief} we recall the basics of the HRT formalism by using a functional formalism and notations that are useful to make the connection with the field-theoretical RG of disordered systems. Section~\ref{sec:QAreplicas} is a summary on the microscopic description of fluids in disordered porous media known as the ``quenched-annealed mixture'' and of the so-called ``replica formalism'' borrowed from spin-glass theory. The next section, section~\ref{sec:naive}, is a first, which after the facts can be dubbed as ``naive'',  attempt to generalize the HRT to the quenched-annealed mixture. We derive the HRT hierarchy and its asymptotic form in the vicinity of the critical point and at long wavelengths. The very same procedure used for the RFIM leads to flow equations that are formally similar to those of the quenched-annealed mixture, giving credit to the equivalence of the critical behaviors of the two systems. We also present a first application in which we analyze the asymptotic RG flow equations for the fluid within an Ornstein-Zernike approximation. The results display inconsistent and unsatisfactory features. Section~\ref{sec:flaws} then is a discussion of the flaws of the naive HRT and a short review of the lessons that can be drawn from a recent nonperturbative RG study of the RFIM. In section~\ref{sec:properHRT}, we lay out the foundations of a proper HRT treatment of fluids in disordered mesoporous materials, deriving the hierarchy of HRT flow equations, their asymptotic form, and proposing approximation schemes. We finally conclude in section~\ref{sec:conclusion} and we provide some additional technical details in two appendices.

\section{HRT in brief}
\label{sec:HRTbrief}

The HRT incorporates in the realm of  liquid-state theory a proper account of the long wavelength fluctuations that play a central role in criticality. This is done by progressively including longer wavelength (shorter wavevector) components of the attractive part of the interactions in the computation of the equilibrium properties of a fluid. In the original HRT formulation\cite{parola-reatto85,HRTreview}, this was achieved by means of a  ``sharp infrared cutoff'' that prevents integration over all wavevectors strictly less than a running value $k$. Since then, it has been reformulated in a more general framework that can accommodate any smooth form of infrared (IR) cutoff procedure\cite{HRTsmooth,caillol09}. To make an easier contact with the field-theoretical RG, most notably that of disordered systems, we use the more compact functional formalism and change notations from the conventional HRT ones; the correspondence with the latter is given in the text and recalled in table \ref{tab:table}.

Consider an atomic  liquid at equilibrium in the presence of an inhomogeneous chemical potential $\mu(x)$, where $x$ represent the coordinates of a point in $d$-dimensional space. The atoms interact via a pair potential that can be decomposed as $v(\vert x-y\vert)=v_R(\vert x-y\vert)+w(\vert x-y\vert)$, where $v_R$ is a steep repulsive interaction (often modeled as a hard core) and $w$ is a longer-ranged attractive interaction that can trigger a liquid-gas transition with a critical point. The grand partition function of the fluid reads
\begin{equation}
\label{eq_Xi_full}
\begin{aligned}
\Xi [\mu]=&\sum_{N}\frac{1}{N!}\int_{x_1}...\int_{x_N} \exp \big\{-\beta \big [V_R(x_1,...,x_N) +\\& \frac{1}{2}  \int_{x}\int_{y} w(\vert x - y\vert) \hat \rho(x) \hat \rho(y) - \int_{x} \mu(x) \hat \rho(x) \big ] \big\},
\end{aligned}
\end{equation}
where $V_R$ is the sum of the pair interactions $v_R$ among all the fluid particles [$V_R=(1/2) \int_{x}\int_{y} v_R(\vert x - y\vert) \hat \rho(x) \hat \rho(y)$], $\hat \rho(x)=\sum_{i=1}^N \delta^{(d)}(x-x_i)$ is the microscopic density of particles at point $x$, $\int_x \equiv \int d^d x$, and $\mu(x)$ may be corrected if necessary by a self-energy term $-w(0)/2$\cite{parola-reatto85,HRTreview}.

The idea behind the exact or nonperturbative RG\cite{wilson,wegner,polshinski,wetterich93,berges02} is to progressively include the effect of fluctuations (more specifically here, the fluctuations generated by the attractive part of the interactions) by integrating over longer and longer wavelengths or, alternatively, smaller and smaller wavevectors. This is achieved by introducing an IR regulator that suppresses the integration over fluctuations of wavevector (momentum)  smaller than a cutoff $k$. At the beginning of the RG flow, $k=\Lambda$ represents a microscopic (``ultra-violet'') scale whose inverse in the present case is of the order of the typical interatomic distance: at this scale, virtually no fluctuations are taken into account, as in a mean-field approximation. On the other hand, when $k=0$, all fluctuations have been included and one obtains the exact theory. 

The HRT combines this RG feature with the requirement of a proper description of the (noncritical) physics mainly induced by the short-ranged repulsive forces in liquids; the IR regulator only modifies the attractive interaction between particles. Let us then define in Fourier space
\begin{equation}
\begin{aligned}
\label{eq_w_k}
w_k(q)= w(q) R_k(q)
\end{aligned}
\end{equation}
where $R_k(q)$ is an IR regulator that goes to zero when $k \rightarrow 0$ and is equal to $1$ when $k=\Lambda$;  by construction then, $w_{k=\Lambda}(q)=w(q)$ and $w_{k=0}(q)=0$. The regulator is also chosen so that it goes to zero as $q\gg k$ and to a $k$-dependent constant (a ``mass'' in field-theoretical language) when $q\ll k$, which will be helpful to limit the contribution of the low-$q$ modes in the grand partition function without affecting that of the high-$q$ modes.

Note the changes of notation with respect to the original HRT literature: the IR cutoff $k$ and the momentum $q$ are denoted by $Q$  and $k$, respectively, in the latter and the $w_k$ defined here corresponds to  $w-w_Q$ in Parola-Reatto's work\cite{parola-reatto85,HRTreview,HRTsmooth}. In addition, we have followed Caillol\cite{caillol09} in the way to introduce the regulator. (Parola and Reatto\cite{HRTsmooth} have a slightly different procedure that uses the functional form of $w(r)$ to define the cutoff.) The reasons for our choice of notations comes from their closer connection to what was done in the field-theoretical nonperturbative RG, more specifically when applied to the theory of disordered systems.

One defines  at the scale $k$ the grand partition function of an equilibrium fluid whose atoms interact through the pair interaction $v_R + w-w_k$ as
\begin{equation}
\label{eq_Xi_k}
\begin{aligned}
\Xi_k [\mu]=&\sum_{N}\frac{1}{N!}\int_{x_1}...\int_{x_N} \exp \big\{-\beta \big [V_R(x_1,...,x_N) +\\& \frac{1}{2} \int_{x}\int_{y} \big[w(\vert x - y\vert) - w_k(\vert x - y\vert)\big]\hat \rho(x) \hat \rho(y)\\& - \int_{x} \mu(x) \hat \rho(x) \big ] \big\},
\end{aligned}
\end{equation}
so that, as a consequence of the properties of the IR regulator, $\Xi_k $ is equal to the grand partition function of the reference system, $\Xi_{k=\Lambda} [\mu]=\Xi_R[\mu]$, at the microscopic scale and is equal to the exact grand partition function of the fully interacting system, $\Xi_{k=0} [\mu]=\Xi[\mu]$ with $\Xi[\mu]$ given in Eq.~(\ref{eq_Xi_full}),  at the end of the flow.

From the grand partition function at the scale $k$, one defines the ``effective average action'' in field-theoretical language\cite{berges02} (which, up to a factor $\beta$, is the Helmholtz free-energy functional in liquid-state theoretical language) through a Legendre transform followed by the addition of a mean-field-like contribution that corrects for the presence of the regulator:
\begin{equation}
\label{eq_Gamma_k}
\begin{aligned}
\Gamma_k[\rho]=& -W_k[\mu]+ \beta  \int_{x} \mu(x) \rho(x)\\&+\frac{\beta}{2}  \int_{x}\int_{y} w_k(\vert x - y\vert)\rho(x)\rho(y),
\end{aligned}
\end{equation}
where $W_k[\mu]=\log \Xi_k [\mu]$ is the generating functional of the connected many-body correlation (or Green's) functions (up to a factor $-\beta$, $W_k$ is a grand-potential functional in liquid-state theoretical language), $\rho(x)=<\hat \rho(x)>$ is the average (inhomogeneous) density field obtained from the Legendre transform as
\begin{equation}
\label{eq_rho_legendre}
\begin{aligned}
\rho(x)= \frac{\delta W_k[\mu]}{\beta \delta \mu(x)}.
\end{aligned}
\end{equation}
(If necessary, one can take account of the self-interaction terms by replacing 
$\mu(x)$ by $\mu(x) -w_k(0)/2$.) The effective average action (modified Helmholtz free-energy functional) $\Gamma_k$ is the generating functional of the ``proper'' or ``$1$-particle irreducible'' (1PI) vertices\cite{zinnjustin89}, which are essentially the direct correlation functions of liquid-state theory. At the microscopic scale $k=\Lambda$, it reduces to a mean-field description of the attractive interactions, 
\begin{equation}
\label{eq_Gamma_Lambda}
\begin{aligned}
\Gamma_{k=\Lambda}[\rho]=\Gamma_R[\rho]+\frac{\beta}{2} \int_{x}\int_{y} w(\vert x- y\vert)\rho(x)\rho(y)
\end{aligned}
\end{equation}
where $\Gamma_R[\rho]$ is the effective action of the reference system. [Note that one could choose $\Lambda \rightarrow \infty$, but the fluctuations generated by the attractive interactions with wavevectors much larger than $2\pi/\sigma$, where $\sigma$ is the typical interatomic distance, essentially do not renormalize eq.~(\ref{eq_Gamma_Lambda}).]  At the macroscopic scale $k=0$, $\Gamma_k$ is equal to the effective action of the fully interacting system,
\begin{equation}
\label{eq_Gamma_full}
\begin{aligned}
\Gamma_{k=0}[\rho]=-\log \Xi[\mu]+ \beta \int_{x} \mu(x) \rho(x).
\end{aligned}
\end{equation}
It is worth stressing that due to the addition of the last term in Eq.~(\ref{eq_Gamma_k}), $\Gamma_k$ is not exactly the Legendre transform of $W_k$ and therefore need not be convex except when $k\rightarrow 0$.

The $2$-point (connected) correlation function is obtained from $W_k[\mu]$ as
\begin{equation}
\label{eq_2point_W_k}
\begin{aligned}
W_k^{(2)}(x,y)= \frac{\delta^2 W_k[\mu]}{\beta \delta \mu(x) \beta \delta  \mu(y)}.
\end{aligned}
\end{equation}
It is related to the standard $2$-point total correlation function $h_k(x,y)$ of the inhomogeneous liquid at scale $k$, in which the atoms interact with the modified pair potential $v_R+ w-w_k$,  according to
\begin{equation}
\label{eq_2pointW_total}
W_k^{(2)}(x,y)=\rho(x) \delta^{(d)}(x-y) +\rho(x)\rho(y) h_k(x,y).
\end{equation}
($W^{(2)}$ is often denoted as $F$ or $F^{(2)}$ in the HRT literature\cite{parola-reatto85,HRTreview,HRTsmooth}.) For a uniform density field $\rho(x)\equiv \rho$, its Fourier transform, $W_k^{(2)}(q,q')=(2\pi)^d\delta^{(d)}(q+q')W_k^{(2)}(q)$, is then connected to the static structure factor $S_k(q)$  through $W_k^{(2)}(q)= \rho S_k(q)$. Similarly, the 2-point ``1PI vertex'' 
\begin{equation}
\label{eq_2point_Gamma_k}
\begin{aligned}
\Gamma_k^{(2)}(x,y)= \frac{\delta^2 \Gamma_k[\rho]}{ \delta \rho(x) \delta  \rho(y)}
\end{aligned}
\end{equation}
is related to the standard Ornstein-Zernike (OZ) direct correlation function $c_k(x,y)$ of the liquid characterized by the modified interaction at scale $k$ according to
\begin{equation}
\label{eq_2pointGamma_direct}
\Gamma_k^{(2)}(x,y)- \beta w_k(\vert x-y\vert)=\frac{\delta^{(d)}(x-y)}{\rho(x)} -c_k(x,y).
\end{equation}
The $2$-point functions $W_k^{(2)}(x,y)$ and $\Gamma_k^{(2)}(x,y)$ are linked through the (modified) Legendre transform as
\begin{equation}
\label{eq_OZ_k}
\begin{aligned}
W_k^{(2)}(x,y)=[\Gamma_k^{(2)}-\beta w_k]^{-1}(x,y),
\end{aligned}
\end{equation}
where the inversion is understood in terms of operators. The above equation when specified to a uniform density (and equivalently uniform chemical potential) is nothing but the familiar OZ equation for the (modified) liquid at scale $k$: $1+\rho h_k(q)=[1-\rho c_k(q)]^{-1}$.

Upon decreasing the infrared cutoff $k$, the effective average action (modified Helmholtz free-energy functional at the scale $k$) $\Gamma_k$ evolves according to the following \textit{exact} ``flow'' equation:
\begin{equation}
\label{eq_ERGE}
\begin{aligned}
\partial_t\Gamma_k[\rho]=\frac{1}{2} \int_{x}\int_{ y} \partial_t \phi_k(\vert x-y\vert) \left [\Gamma_{k}^{(2)}[\rho]+ \phi_k \right ]^{-1}(x,y),
\end{aligned}
\end{equation}
where $t=\log(k/\Lambda)$ and we have (conventionally) introduced the notation $\phi_k=-\beta w_k$, which is a positive quantity as $w_k$ is an attractive interaction. By differentiating Eq.~(\ref{eq_ERGE}) with respect to the density field and evaluating the resulting expressions for a uniform density field $\rho(x)=\rho$, one generates an exact  hierarchy of coupled flow equations for the 1PI vertices (modified direct correlation functions), which is the more familiar form of the HRT. For instance, the first equation for the Helmholtz free-energy density $\mathcal A_k(\rho) =\Gamma_k(\rho)/V$ (or ``effective average potential'' in field-theoretical language) reads
\begin{equation}
\label{eq_ERGE_unif}
\begin{aligned}
\partial_t \mathcal A_k(\rho)=\frac{1}{2} \int_{q}\frac{\rho\, \partial_t \phi_k(q)}{1-\rho c_k(q;\rho)},
\end{aligned}
\end{equation}
where $\int_q \equiv d^q/(2\pi)^d$ and $c_k(q;\rho)$ is the direct OZ correlation function that is related to the $2$-point 1PI vertex by $1/\rho - c_k(q;\rho) = \Gamma_{k}^{(2)}(q;\rho)+\phi_k(q)$ [see Eq.~(\ref{eq_2pointGamma_direct})]. Note that we have defined $\mathcal A_k$ as the opposite of the definition chosen by Parola and Reatto\cite{parola-reatto85,HRTreview,HRTsmooth}: in our notation $\mathcal A_k$ is directly $\beta$ times the modified Helmholtz free-energy density at scale $k$.

\begin{table*}[htbp]
\begin{center}
\begin{tabular}{|c|c|}
\hline
\ \ \ \ $field-theoretical\,RG$ \ \ \ \ & \ \ \ \ $HRT$ \ \ \ \ \\
\hline
\hline
\mbox{IR cutoff} $k$ &  \mbox{IR cutoff} $Q$ \\
\hline
\mbox{Momentum} $q$ &  \mbox{wavevector} $k$\\
\hline
\mbox{Effective average action} $\Gamma_k$ & \mbox{modified Helmholtz free-energy functional } \\
\hline
\mbox{Effective average potential} $\Gamma_k(\phi)/V=\mathcal A_k$& \mbox{modified Helmholtz free-energy density} $-\mathcal A_Q(\rho)$ \\
\hline
\mbox{Proper (1PI) vertices: at the 2-point level,} $\Gamma_k^{(2)}$ &\mbox{Modified direct correlation functions: at the pair level,} $\mathcal C_Q$\\
\hline
\mbox{Green's (correlation) functions: at the 2-point level,} $W_k^{(2)}$ &\mbox{(Connected) correlation functions: at the pair level,} $F_Q$\\
\hline
\mbox{Regulator for fluctuations:} $\phi_k(q)=-\beta R_k(q)w(q)$ & \mbox{Modified interaction:} $\phi(k)-\phi_Q(k)= - \beta [w(k)- w_Q(k)]$\\
\hline
\mbox{Legendre transform relation:} $W_k^{(2)}=(\Gamma_k^{(2)}+\phi_k)^{-1}$ & \mbox{Ornstein-Zernike equation:} $F_Q=(-\mathcal C_Q+\phi-\phi_Q)^{-1}$\\
\hline
\end{tabular}
\end{center}
\caption{Table of equivalence between the field-theoretical nonperturbative RG and the HRT for fluids.}
\label{tab:table}
\end{table*}

The above functional integro-differential equation for the RG flow of the effective average action, Eq.~(\ref{eq_ERGE}), is exact but of course impossible to solve in general. Approximation schemes must be devised. In field theory, nonperturbative approximations are formulated directly at the level of the generating functional\cite{wetterich93,berges02}: an ansatz is chosen for the effective average action and consistency is automatically enforced for the relations between the derivatives of the latter and the 1PI vertices. In the HRT of fluids, it is not as easy to work at the level of the generating functional, here the Helmholtz free-energy functional; approximations are proposed as truncations of the hierarchy of coupled equations for the 1PI vertices for a uniform density field. The common implementation corresponds to formulating an ansatz for the $2$-point 1PI vertex function, here the direct pair correlation function. One has then to be careful to properly enforce the sum rules that relate the different levels of the hierarchy. For instance, the $2$-point direct correlation function must satisfy the ``compressibility sum rule'' that relates it to the Helmholtz free energy, namely,
\begin{equation}
\label{eq_compress_sum_rule}
\begin{aligned}
\frac{1}{\rho}- c_k(q=0;\rho)-\phi_k(q=0)=\Gamma_{k}^{(2)}(q=0;\rho)=\frac{\partial^2\mathcal A_k(\rho)}{\partial \rho^2},
\end{aligned}
\end{equation}
where we have used the definitions and notations introduced above.

A typical closure of the HRT hierarchy can be written as\cite{parola-reatto85,HRTreview,HRTsmooth,HRTapplic,HRTkahl02,HRTverso06} 
\begin{equation}
\label{eq_closure}
\begin{aligned}
\frac{1}{\rho}- c_k(q;\rho)-\phi_k(q)=\frac{\partial^2\mathcal A_k(\rho)}{\partial \rho^2}+ \delta \Sigma_k(q;\rho),
\end{aligned}
\end{equation}
where an approximation is formulated for $\delta \Sigma_k(q;\rho)$ that satisfies $\delta \Sigma_k(q=0;\rho)=0$. Actually, the theory being regularized for wavevectors $q\lesssim k$, it is expected that a relatively simple ansatz for $c_k(q;\rho)$, or equivalently for $\delta \Sigma_k(q;\rho)$, already captures most of the physics. All of the proposed approximations have been of ``OZ type'' as they assume that the $q$-dependence of $\delta \Sigma_k(q;\rho)$ is analytic at small $q$'s and therefore starts as $q^2$. The simplest nontrivial such approximations are of RPA form with \textit{e.g.}
\begin{equation}
\label{eq_RPA}
\begin{aligned}
\delta \Sigma_k(q;\rho)=&-[c_R(q;\rho)-c_R(q=0;\rho)]\\&- \left [\phi(q)-\phi(q=0) \right ],
\end{aligned}
\end{equation}
or\cite{HRTapplic}
\begin{equation}
\label{eq_RPAbis}
\begin{aligned}
\delta& \Sigma_k(q;\rho)=-[c_R(q;\rho)-c_R(q=0;\rho)]\\&+\left[\frac{\phi(q)-\phi(q=0)}{\phi(q=0)}\right ]\left[\frac{\partial^2\mathcal A_k(\rho)}{\partial \rho^2} + c_R(q=0;\rho)-\frac{1}{\rho}\right ].
\end{aligned}
\end{equation}
The latter expression comes from the OZ-like closure $c_k(q;\rho)+\phi_k(q)=c_R(q;\rho)+\lambda_k(\rho)\phi(q)$, where the $T$ and $\rho$ dependent factor $\lambda_k$ is adjusted so that Eq.~(\ref{eq_compress_sum_rule}) is satisfied. Improved approximations based on ORPA, MSA, etc,  which allow one to enforce the ``core'' condition when the reference pair potential is a hard-sphere one, have also been implemented\cite{parola-reatto85,HRTreview,HRTsmooth,HRTapplic,HRTkahl02,HRTverso06}.

The main drawback of all these OZ-like approximations is that they do not properly describe the singular wave-vector dependence of the correlation function at the critical point, \textit{i.e.} the fact that when $k=0$, $\delta \Sigma_k(q)$ starts as $q^{2-\eta}$ instead of $q^2$, where $\eta>0$ is a critical exponent called the ``anomalous dimension'' of the field. [$\delta \Sigma_k(q)$ is analytic  for $q \lesssim k$ but the singularity emerges when $q \sim k \rightarrow 0$.] This weakness is not too significant when describing the critical behavior of pure fluids or mixtures, as $\eta\simeq0.03$ is then very small, but it is much more serious when considering the critical behavior of a fluid in a disordered porous medium. This will be further discussed below.

Finally, we close this brief review on the HRT by considering the sharp cutoff originally proposed. In this case, the regulator is then a Heaviside step function, $R_k(q)\propto H(k-q)$, so that $w_k=0$ when $q>k$ and $w_k = w$ when $q<k$. Its derivative $\partial_t R_k(q)$, which appears in the exact flows equation Eq.~(\ref{eq_ERGE}), is then proportional to a delta function, $\delta(k-q)$. Some care is needed when handling the ``loop integral'' appearing in the right-hand side of the flow equations as it involves a combination of delta and Heaviside functions [see for instance Eq.~(\ref{eq_ERGE_unif})]. This point is detailed in Refs.~[\onlinecite{morris-sharp}, \onlinecite{caillol09}]. After a proper account of this problem,  Eq.~(\ref{eq_ERGE_unif}) can be rewritten as
\begin{equation}
\label{eq_ERGE_unif_sharp}
\begin{aligned}
\partial_t \mathcal A_k(\rho)=\frac{1}{2} \, k^{d}\, v_d \log \left (1+\frac{\phi(k)}{\Gamma_k^{(2)}(k;\rho)}\right ),
\end{aligned}
\end{equation}
where $v_d^{-1}=2^{d+1}\pi^{d/2}\Gamma(d/2)$ and $\phi(k)=-\beta w(q=k)$. The above flow equation is identical to that derived by Parola and Reatto (with the change of notations $Q \rightarrow k$, $\mathcal A_Q \rightarrow -\mathcal A_k$, $\mathcal C_Q \rightarrow \Gamma_k^{(2)}$, and $\phi_Q \rightarrow \phi-\phi_k$). The same is true for the higher orders of the HRT hierarchy.

\section{Quenched-annealed mixtures, replica formalism, and liquid-state theory}
\label{sec:QAreplicas}

The first step in the description of the  behavior of fluids adsorbed in disordered mesoporous materials is to provide a realistic model of the latter. The approach that has been used in the past two decades, both in computer simulations and in statistical-mechanical theories\cite{review-gubbins,review-mlr}, is to treat the solid phase of the porous material (the ``matrix'') as a collection of particles in some predefined microstructure which is assumed to be rigid\cite{madden88,QAgiven-stell}. Therefore, the solid is simply viewed as a configuration of quenched particles, sampled from a given probability distribution. This distribution can be obtained either by modelling the process of formation of the porous material and considering the resulting configurations (as done for instance for the porous glass Vycor formed by spinodal decomposition\cite{review-gubbins} or for the silica aerogel modelled by a cluster-cluster aggregation\cite{jullien,QAkrako}) or by taking some chosen equilibrium distribution of particles (hard spheres or ideal gas).

The fluid establishes itself in a state of thermal equilibrium in the presence of, and in interaction with, this quenched structure. Such a ``quenched-annealed mixture''\cite{madden88,QAgiven-stell} is macroscopically homogeneous, but the external field exerted by the matrix on the fluid varies on micro- and meso-scopic scales, thus making the fluid strongly inhomogeneous on such scales. This hampers the direct application of many common methods of liquid-state theory. For instance, density-functional schemes, which have proven very useful for fluids in the presence of a substrate of simple geometry\cite{evans},  are intractable in general for a realistic disordered porous medium (the required numerical resolution becomes only barely possible for simplified lattice-gas models\cite{kierk-hysteresis,detche-hysteresis,DFTkierlik01}).

It was shown\cite{QAgiven-stell,QArosinberg} that the difficulty can be, at least formally, circumvented by applying the ``replica method'' developed in the theory of spin glasses\cite{SGbook}. The study of the original quenched-annealed mixture is then replaced by that of a (fictitious) homogeneous mixture of $n+1$ components at equilibrium: one component is formed by the matrix particles now considered as fully annealed, and the others are $n$ identical replicas of the fluid. The matrix interacts with all fluid replicas, but the latter do not interact with each other. For this $(n+1)$-component mixture, one can use the whole machinery of liquid-state statistical mechanics and, at the end of the manipulations and calculations, take the $n \rightarrow 0$ limit with an appropriate analytic continuation.

Consider for instance the grand potential $\Omega_f(\mu)$ of the fluid in equilibrium at a chemical potential $\mu$ within a porous material [in the field-theoretical language used in the above section, $-\beta \Omega_f(\mu)=W_f(\mu)$]. It is obtained as
\begin{equation}
\label{eq_Omega_fluid_QA}
\begin{aligned}
-\beta  \Omega_f(\mu)&= \int_{y_1}...\int_{y_{N_0}} \mathcal P_0(\{y_i\}_{N_0})\log \Xi_f(\mu;\{y_i\}_{N_0})\\&= \overline{\log \Xi_f(\mu;\{y_i\}_{N_0})} ,
\end{aligned}
\end{equation}
where $\{y_i\}_{N_0}$ denotes a configuration of $N_0$ matrix particles, $\mathcal P_0(\{y_i\}_{N_0})$ is the probability of finding this configuration, and the overline in the second line is a compact way to express the average over the ``quenched disorder'', \textit{i.e.}, the frozen-matrix configurations; $\Xi_f(\mu;\{y_i\}_{N_0})$ is the matrix-dependent grand partition function which is defined as
\begin{equation}
\label{eq_Xi_matrixdepdt}
\begin{aligned}
\Xi_f(\mu;\{y_i\}_{N_0})& = \sum_{N} \frac{1}{N!} \int_{x_{1}}...\int_{x_{N}} \exp \big\{-\beta \big[ V_{ff}(\{x_{i}\}_N)\\& +V_{mf}(\{x_{i}\}_N ;\{y_i\}_{N_0})-\int_x \mu(x)\hat  \rho(x) \big ] \big\},
\end{aligned}
\end{equation}
where  $V_{ff}(\{x_{i}\}_N)$ and $V_{mf}(\{x_{i}\}_N ;\{y_i\}_{N_0})$ are the total fluid-fluid interaction potential and the matrix-fluid interaction potential, respectively, and $\hat  \rho(x)$ is the microscopic density of fluid particles at point $x$. The Helmholtz free energy of the fluid (or effective action in field-theoretical language) is as usual the Legendre transform of the grand potential  $\Omega_f(\mu)$.

To perform the average over the matrix configurations of the logarithm in Eq.~(\ref{eq_Omega_fluid_QA}), one can use the replica trick\cite{SGbook} which simply states that
\begin{equation}
\label{eq_Omega_replicatrick}
\begin{aligned}
\beta \Omega_f(\mu)&= \lim_{n\rightarrow 0} \frac{1}{ n} \overline{\big (- \exp [n \log \Xi_f(\mu;\{y_i\}_{N_0}]+1\big )}\\&=\lim_{n\rightarrow 0} \frac{1}{ n}\left (1-\exp[-\beta \Omega_f^{rep}(\mu;N_0)]\right )\\&=\lim_{n\rightarrow 0}\left [\frac{\beta \Omega_f^{rep}(\mu;N_0)}{n}\right ],
\end{aligned}
\end{equation}
where
\begin{equation}
\label{eq_Omega_replicated}
\begin{aligned}
&-\beta \Omega_f^{rep}(\mu;N_0)= \log \int_{y_1}...\int_{y_{N_0}} \sum_{N_1,...,N_n}\frac{1}{N_1!...N_n!}\\&\int_{\{x_{1i}\}_{N_1}}...\int_{\{x_{ni}\}_{N_n}}\mathcal P_0(\{y_i\}_{N_0}) \exp \big\{-\beta  \sum_{a=1}\big[V_{ff}(\{x_{ai}\}_{N_a})\\& +V_{mf}(\{x_{ai}\}_{N_a} ;\{y_i\}_{N_0})-\int_x \mu(x) \hat \rho_a(x)\big ] \big\},
\end{aligned}
\end{equation}
with $\hat \rho_a(x)$  the microscopic fluid density of replica $a$. $\Omega_f^{rep}(\mu;N_0)$ corresponds to an \textit{equilibrium} thermodynamic potential for a mixture of $n$ fluid replicas at the chemical potential $\mu$ and one matrix species at fixed number of particle $N_0$ (and it is proportional to the number $n$ of replicas). Thanks to the replica trick, the quenched-annealed binary mixture has been replaced by a fully annealed mixture of $n+1$ species. The fluid particles do not interact directly when they belong to distinct replicas but they all interact with the matrix particles. In the following, the (now annealed) matrix is characterized by a subscript $0$ and the fluid replicas take the indices $1,2,...,n$. We denote by Greek letters, $\alpha, \beta,...$, the $n+1$ species (\textit{e.g.}, $\alpha=0,1,...,n$) and by Roman letters, $a,b,...$, the $n$ fluid replicas (\textit{e.g.}, $a=1,...,n$).

One can then apply all the tools of liquid-state theory to this $(n+1)$-component equilibrium mixture. In particular, one can introduce direct and total correlation functions (or in field-theoretical language, proper or 1PI vertices and Green's functions) and relate them via OZ equations that stem from the Legendre transform between grand potential (generating functional of the Green's functions) and Helmholtz free energy (effective action, which is the generating functional of the 1PI vertices). At the pair level and using the field-theoretical notations (see section~\ref{sec:HRTbrief}), one has in Fourier space and in matrix form
\begin{equation}
\label{eq_OZfield_replicas}
\begin{aligned}
\vect W^{(2)}(q)=\vect \Gamma^{(2)}(q)^{-1}
\end{aligned}
\end{equation}
or equivalently in liquid-state notations,
\begin{equation}
\label{eq_OZliquid_replicas}
\begin{aligned}
\vect \rho + \vect \rho \vect h(q) \vect \rho=[\vect \rho^{-1}-\vect c(q)]^{-1},
\end{aligned}
\end{equation}
where the matrices are $(n+1) \times (n+1)$ and $\vect \rho$ is diagonal with $\rho_{00}=\rho_0$ and $\rho_{11}=...=\rho_{nn}=\rho$.

To recover the properties of the original system, an analytic continuation to arbitrary values of $n$ together with the limit $n \rightarrow 0$ have to be taken. Assuming that the permutational symmetry among the $n$ replicas is not broken (contrary to what occurs for instance for the Sherrington-Kirkpatrick spin-glass model\cite{SGbook}), one can easily diagonalize the $(n+1)\times (n+1)$ matrices and one ends up in the limit $n\rightarrow 0$ with the following ``replica-symmetric OZ equations''\cite{QAgiven-stell}:
\begin{equation}
\label{eq_ornstein_zernike_QA_RS}
\begin{aligned}
&h_{mf}(q)=S_{mm}(q)\frac {c_{mf}(q)}{1-\rho c_{con}(q)}\, ,\\&
1+\rho h_{con}(q)=\frac {1} {1-\rho c_{con}(q)}\, ,\\&
h_{dis}(q)=\frac {c_{dis}(q)+\rho_0 S_{mm}(q)c_{mf}(q)^2} {[1-\rho c_{con}(q)]^2}\, .
\end{aligned}
\end{equation}
In the above equation,  $S_{mm}(q)$ is the structure factor of the matrix phase of the porous material, $h_{con}$ and $h_{dis}$ are the so-called ``connected'' and ``disconnected'' (or ``blocking''\cite{QAgiven-stell}) fluid-fluid pair correlation functions. In physical terms, the latter are defined from the $1$-body and $2$-body densities $\rho_{f}(x)$ and $\rho_{ff}(x,y)$ of the fluid in a given configuration of the solid matrix as
\begin{equation}
\label{eq_hconn_hdisc}
\begin{aligned}
&\rho^2 h_{con}(\vert x_{1}-x_2\vert)=\overline{\rho_{ff}(x_1,x_2)-\rho_{f}(x_1)\rho_{f}(x_2)}\\&
\rho^2 h_{dis}(\vert x_{1}-x_2\vert)=\overline{\rho_{f}(x_1)\rho_{f}(x_2)}-\rho^2,
\end{aligned}
\end{equation}
and their sum is equal to the total fluid-fluid pair correlation function $h_{ff}$ that is directly related to the fluid-fluid structure factor through $S_{ff}(q)=1+\rho h_{ff}(q)$; finally, $\rho=\overline{\rho_{f}}$ is the mean fluid density.

In terms of the replicated mixture, one has
\begin{equation}
\label{eq_hconn_hdisc_RS}
\begin{aligned}
&h_{con}(q)=\lim_{n\rightarrow 0} h_{aa}(q) - h_{dis}(q)\\&
h_{dis}(q)=\lim_{n\rightarrow 0} h_{ab}(q),
\end{aligned}
\end{equation}
where $a \neq b$ in the last equation. Similarly, the static structure factor of the matrix can be expressed as $S_{mm}(q)=1+\rho_0 \lim_{n\rightarrow 0} h_{00}(q)$ and the matrix-fluid correlation function as $h_{mf}(q)=\lim_{n\rightarrow 0} h_{0a}(q)=\lim_{n\rightarrow 0} h_{a0}(q)$.

The connected fluid-fluid pair correlation functions and the average Helmholtz free-energy density for the fluid in the porous material $A_f(\rho;\rho_0)$ are related by a compressibility sum rule\cite{QArosinberg},
\begin{equation}
\label{eq_compress_connQA_RS}
\begin{aligned}
\frac{1}{\rho}- c_{con}(q=0;\rho)&=\frac{1}{\rho+\rho^2 h_{con}(q=0;\rho)}\\&= \frac{\partial^2 \beta A_f(\rho;\rho_0)}{\partial \rho^2}.
\end{aligned}
\end{equation}
A similar relation holds for the matrix-fluid direct correlation function:
\begin{equation}
\label{eq_compress_mfQA_RS}
\begin{aligned}
- c_{mf}(q=0;\rho)= \frac{\partial^2 \beta A_f(\rho;\rho_0)}{\partial \rho \partial \rho_0}.
\end{aligned}
\end{equation}
However, neither the total nor the direct disconnected fluid-fluid correlation functions are related to derivatives of the average Helmholtz free-energy density for the fluid $A_f(\rho)$.

Over the last two decades, there has been a great deal of activity to solve the replica-symmetric OZ equations with various standard closures\cite{hansen-mcdo} (Percus-Yevick, HNC, MSA, ORPA, EXP and OCT, etc...) and various choices of microstructure for the solid matrix [which enters the equations only through the structure factor $S_{mm}(q)$]. Fluid phase diagrams and pair correlation functions have been computed and compare well, at least semi-quantitatively, with the existing simulation and experimental data\cite{QAkrako,QAglandt,QAlomba,QAmonson,QAkahl01,QAkierlik-monson02}. However, the standard approximations break down near the critical points, when present, because of their inability to describe correctly the long-range fluctuations. A promising route is then to apply the HRT to the quenched-annealed mixtures.

\section{A naive HRT approach}
\label{sec:naive}

\subsection{Strategy}

The strategy that appears natural to describe the critical behavior of fluids in a disordered porous medium consists in generalizing the HRT approach summarized in section~\ref{sec:HRTbrief} to the replicated mixture with $n+1$ components introduced in section~\ref{sec:QAreplicas}. We choose for reference system the mixture defined with all interaction pair potentials (matrix-fluid and fluid-fluid), except the attractive interaction $w_{ff}$ between fluid particles in the same replica (recall that there are no direct interactions between fluid particles belonging to different replicas). Generalizing Eq.~(\ref{eq_w_k}), we therefore introduce a modified interaction matrix $w_{k;\alpha \beta}$ in which the only nonzero elements are the diagonal terms $w_{k;aa}$ with $a=1,...,n$, which are defined such that
\begin{equation}
\begin{aligned}
\label{eq_w_kaa}
w_{k;aa}(q)= R_k(q)w_{ff}(q) ,
\end{aligned}
\end{equation}
with $R_k(q)$ having the same properties as for a one-component fluid (see section~\ref{sec:HRTbrief}). As before, we also introduce $\phi_k(q)=-\beta R_k(q) w_{ff}(q)$.

With the above choice of reference system and IR regulator, the extension of the exact functional HRT equation to the equilibrium $(n+1)$-component mixture reads
\begin{equation}
\label{eq_ERGE_replicated_naive}
\begin{aligned}
&\partial_t\Gamma_{k}^{rep}[\{\rho_{\alpha}\}]=\\&\frac{1}{2} \int_{q} \partial_t \phi_k(q) \sum_{a=1}^n\bigg (\left [\vect \Gamma_{k}^{rep(2)}[\{\rho_{\alpha}\}]+ \phi_k \vect I \right ]^{-1}\bigg)_{aa}(q,-q)
\end{aligned}
\end{equation}
where we have momentarily introduced inhomogeneous density fields $\rho_a(x)$ that are different for the different replicas; $\vect \Gamma_{k}^{rep(2)}$ and $\vect 1$ are $(n+1)\times (n+1)$ matrices with $\Gamma_{k;\alpha \beta}^{rep(2)}(x,y)=\delta^2\Gamma_{k}^{rep}/\delta \rho_{\alpha}(x)\delta \rho_{\beta}(y)$ and $I_{\alpha \beta}=\delta_{\alpha \beta}(1-\delta_{\alpha 0})$. As stressed already several times, the effective average action $\Gamma_{k}^{rep}$ is, up to a factor $\beta$, the Helmholtz free-energy functional and the $2$-point 1PI vertices $\Gamma_{k;\alpha \beta}^{rep(2)}$ are related to the direct pair correlation functions [see Eqs.~(\ref{eq_OZfield_replicas}) and (\ref{eq_OZliquid_replicas})]. Another more compact way to write the HRT equation is by introducing an operator $\tilde \partial_t$ that acts \textit{only} on the $k$-dependence of $\phi_k(q)$. This leads to
\begin{equation}
\label{eq_ERGE_replicated_tilde}
\begin{aligned}
&\partial_t\Gamma_{k}^{rep}[\{\rho_{\alpha}\}]=\\&\frac{1}{2}  \tilde \partial_t  \int_{q}  \, tr \big( \log \left [\vect \Gamma_{k}^{rep(2)}[\{\rho_{\alpha}\}]+ \phi_k \vect I \right ]\big)(q,-q),
\end{aligned}
\end{equation}
where $tr$ is a trace over the $n+1$ components. The whole HRT hierarchy for 1PI vertices (direct correlation functions) evaluated for uniform density fields is again simply obtained from the above functional equation by repeated differentiations.

To put this formal scheme at work, we have to (i) consider an analytic continuation in the number $n$ of replicas and take the limit $n \rightarrow 0$,  (ii) devise an approximation to truncate the HRT hierarchy, and (iii) choose a specific cutoff function $R_k(q)$.

\subsection{Replica-symmetric HRT}

Concerning the first point mentioned above [point (i)], we make the assumption, which is common in the treatment of quenched-annealed mixtures (see section~\ref{sec:QAreplicas}), that the permutational symmetry between replicas is not broken when $n \rightarrow 0$ (provided all the densities $\rho_a$ are taken as equal, which they should be when describing the physical system). This implies in particular that any $(n+1)\times (n+1)$ matrix $\vect M$ appearing in the HRT hierarchy has the same structure, with $M_{0a}=M_{01}$ and $M_{ab}=(M_{11}-M_{12})\delta_{ab} + M_{12}$, for $a,b=1,...,n$. Such a matrix can be easily diagonalized, which henceforth allows for straightforward algebraic manipulations. For instance, one simply obtains
\begin{equation}
\label{eq_log_replicas}
\begin{aligned}
tr \log \vect M=&\log(M_{00}) +n \bigg [\log(M_{11}-M_{12})+\frac{M_{12}-\frac{M^2_{01}}{M_{00}}}{M_{11}-M_{12}}\bigg ]\\& +O(n^2).
\end{aligned}
\end{equation}
Similar but more complicated properties hold for the higher-order tensors, \textit{e.g.}, $\Gamma_{k;\alpha \beta \gamma}^{rep(3)}(x,y,z)$.

The first equation of the HRT hierarchy in the $n \rightarrow 0$ limit then provides a flow equation for the (modified) Helmholtz free-energy density of the fluid at scale $k$, $\mathcal A_{f,k}(\rho;\rho_0)=\lim_{n \rightarrow 0}\partial [\Gamma_{k}^{rep}(\rho;\rho_0)/V]/\partial n$, which reads
\begin{equation}
\label{eq_ERGE_A_k_naive_RS}
\begin{aligned}
\partial_t\mathcal A_{f,k}(\rho;\rho_0)=&\frac{1}{2} \int_{q} \partial_t \phi_k(q) \bigg \{ \frac{1}{\rho^{-1}-c_{con,k}(q)} \\& + \frac{c_{dis,k}(q)+\rho_0 S_{mm}(q)c_{mf,k}(q)^2 }{[\rho^{-1}-c_{con,k}(q)]^2}\bigg\},
\end{aligned}
\end{equation}
where the direct correlation functions at the scale $k$ depend on $\rho$ and $\rho_0$ and are related to the 1PI vertices according to
\begin{equation}
\label{eq_direct_1PI_replicas}
\begin{aligned}
&\rho^{-1}-c_{con,k}(q) = \Gamma_{k;11}^{rep(2)}(q)-\Gamma_{k;12}^{rep(2)}(q)+ \phi_k(q),\\&
c_{dis,k}(q) = -\Gamma_{k;12}^{rep(2)}(q),\\&
c_{mf,k}(q) = - \Gamma_{k;01}^{rep(2)}(q)
\end{aligned}
\end{equation}
in the limit $n \rightarrow 0$. The first term of the right-hand side of Eq.~(\ref{eq_ERGE_A_k_naive_RS}) is similar to that in the HRT for the bulk fluid [see Eq.~(\ref{eq_ERGE_unif})] and the second term is really the signature of the presence of the disordered porous matrix: it vanishes identically when $\rho_0=0$ while, then, $c_{con,k}\equiv c_k$. Note that the total pair correlation functions are related to the above direct correlation functions through an obvious generalization at scale $k$ of the replica-symmetric OZ equations, Eq.~(\ref{eq_ornstein_zernike_QA_RS}).

The other equations of the hierarchy can be written in the same way as Eq.~(\ref{eq_ERGE_A_k_naive_RS}). The derivation is quite straightforward, but, because of the rapidly increasing number of irreducible components of the many-body correlation functions that one must consider when dealing with higher orders, the explicit expressions are too lengthy to be reproduced here, even in a diagrammatic representation. (An illustration will be given below for the flow equation of the connected pair correlation function in the asymptotic regime near the critical point.) Note that there are no flow equations for the pure matrix functions in the limit $n \rightarrow 0$, which is due to the fact that the matrix is rigid and independent of the fluid.

Exact compressibility-like sum rules are also obtained, that relate the connected component of the fluid-fluid direct correlation function and the matrix-fluid direct correlation function at zero wavevector to the second derivatives of  $\mathcal A_{f,k}$:
\begin{equation}
\label{eq_compress_sum_rule_con}
\begin{aligned}
\frac{1}{\rho}- c_{con,k}(q=0)-\phi_k(q=0)=\frac{\partial^2\mathcal A_{f,k}(\rho;\rho_0)}{\partial \rho^2},
\end{aligned}
\end{equation}
\begin{equation}
\label{eq_compress_sum_rule_mf}
\begin{aligned}
- c_{mf,k}(q=0)=\frac{\partial^2\mathcal A_{f,k}(\rho;\rho_0)}{\partial \rho \partial \rho_0}.
\end{aligned}
\end{equation}
Generalizations of these identities can be derived for the higher orders.

\subsection{Asymptotic behavior}
\label{sub:asymptotic}

When approaching the gas-liquid critical point (in the cases of course where the effect exerted by the porous material is not so strong as to destroy the gas-liquid transition of the adsorbed fluid), the fluid compressibility diverges. Accordingly, the total connected fluid-fluid correlation function at the critical point behaves as
\begin{equation}
\label{eq_correl_con_crit}
\begin{aligned}
h_{con,crit}(q) \sim q^{-(2-\eta)}.
\end{aligned}
\end{equation}
at small wavevector $q$. The asymptotic regime of the HRT equations describes the long-distance physics near the critical point, which corresponds to small IR cutoff $k$ and small wavevectors. Keeping in mind the definition of the total and direct correlation functions,  the above critical behavior translates into the following small $k$ behavior of the connected pair correlation functions:
\begin{equation}
\label{eq_vertex2conn_asymptotic_naive}
\begin{aligned}
\frac{1}{\rho}-c_{con,k}(q)-\phi_k(q) \simeq k^{2-\eta}\, u_{con,k}(\hat q^2),
\end{aligned}
\end{equation}
where $\hat q^2=q^2/k^2$ and the scaling function $u_{con,k}$ is expected to converge to a finite limit $u_{con,*}$ when $k\rightarrow 0$. To be consistent with this behavior, we choose the IR regulator such that $\phi_k(q)\simeq k^{2-\eta}$; more specifically, we define
\begin{equation}
\label{regulator}
R_k(q)=Z_k k^2 r(\frac{q^2}{k^2}),
\end{equation}
where the function $r(y)$ is equal to $1$ when $y=0$ and decreases fast enough to zero when $y\rightarrow \infty$ [to cite a few examples used in the nonperturbative-RG literature: $(1-y)H(1-y)$\cite{litim}, $y/(\exp(y)-1)$\cite{wetterich93,berges02}, or the sharp cutoff $H(1-y)$, where as before $H$ is the Heaviside step function]. $Z_k \sim k^{-\eta}$ is introduced to account for the so-called field renormalization (which introduces the ``anomalous dimension'' $\eta$) and is defined through a prescription on the direct pair correlation function (see below).   In the asymptotic regime, for wavevectors such that $q/k=\hat q = O(1)$, the cutoff interaction $\phi_k(q)$ can then be replaced by
\begin{equation}
\label{eq_phi_k_asympt}
\phi_k(q)\simeq -\beta w_{ff}(q=0) Z_k k^2 r(\hat q^2),
\end{equation}
where we recall that $w_{ff}(0)$ is negative. One then finds, after dropping some irrelevant constants that can always be incorporated in a redefinition of the functions and variables, that $\rho^{-1}-c_{con,k}(q)\sim k^{2-\eta}[u_{con,k}(\hat q^2)+r(\hat q^2)]$, which gives back Eq.~(\ref{eq_correl_con_crit}) (provided again that $\lim_{k\rightarrow 0} u_{con,k}(\hat q^2)$ is finite and different from zero).

On the other hand, at the critical point, the matrix-fluid direct correlation function at zero wavevector satisfies
\begin{equation}
\label{eq_direct_mf_crit}
\begin{aligned}
c_{mf,crit}(q=0)=-\frac{\partial^2 \mathcal A_f}{\partial \rho \partial \rho_0}\bigg \vert_{crit}=-\frac{\partial \mu_{crit}}{\partial \rho_0},
\end{aligned}
\end{equation}
which is finite. Therefore, in the asymptotic regime, $\rho_0 S_{mm}(q) c_{mf,k}(q)^2$ is finite. The fluid-fluid disconnected total pair correlation function $h_{dis,k}(q)$, which is given in Eq.~(\ref{eq_ornstein_zernike_QA_RS}), has then two possible types of behavior. If an accidental cancellation occurs such that $c_{dis,k}(q)+\rho_0 S_{mm}(q)c_{mf,k}(q)^2\rightarrow 0$, then the two terms of the right-hand side of Eq.~(\ref{eq_ERGE_A_k_naive_RS}) could possibly scale in the same way if $c_{dis,k}(q)+\rho_0 S_{mm}(q)c_{mf,k}(q)^2$ goes to zero as $k^{2-\eta}$. This is actually the case when a special symmetry is present that leads to a cancellation of the random chemical potentials exerted by the solid matrix on the fluid. For a lattice-gas version of the quenched-annealed mixture, we showed  that when the ratio between the attractive matrix-fluid interaction and the fluid-fluid one is exactly equal to $2$, the system is isomorphic to the so-called site-diluted Ising model\cite{SDIMkierliketal}. In such a case, the symmetries of the Hamiltonian impose that, for the fully interacting system (\textit{i.e.}, $k=0$) along the critical isochore,  $c_{dis,k=0}(q)\equiv 0$  whereas $c_{mf,k=0}(q)\equiv (1/2)[-\rho^{-1}+c_{con,k=0}(q)]$ goes to zero as one approaches the critical point\cite{SDIMkierliketal}. As a consequence, $c_{dis,k}(q)+\rho_0 S_{mm}(q)c_{mf,k}(q)^2\rightarrow 0$ as $1-\rho c_{con,k}(q)$ when $k\rightarrow 0$  in the critical region. The site-diluted Ising model has an upper critical dimension of $4$ and quenched disorder plays a much weaker role there than in the RFIM. It should thus be handled in a specific manner (in particular,  if one wishes to preserve the symmetry property along the flow, one needs to make the IR regulator also act on the matrix-fluid attractive interaction) and will not be further considered in this paper.

In the generic case, one expects $c_{dis,k}(q)+\rho_0 S_{mm}(q)c_{mf,k}(q)^2\neq 0$. The second term of the right-hand side of Eq.~(\ref{eq_ERGE_A_k_naive_RS}) then asymptotically dominates the first term at small $k$. As a result, the latter is irrelevant and can be dropped when $k\rightarrow 0$. The quenched disorder then controls the critical behavior. Technically, this corresponds to a ``zero-temperature fixed point'' which is  characterized by an additional exponent $\theta$ and a violation of the so-called hyperscaling relation,
\begin{equation}
\label{eq_hyperscaling}
2-\alpha=\nu(d-\theta),
\end{equation}
where $\alpha$ and $\nu$ are the specific-heat and correlation-length critical exponents. Such a zero-temperature fixed point is known to control the critical behavior of the RFIM\cite{villain84,fisher86,middleton-fisher02}.

One then anticipates that the disconnected  component asymptotically behaves as
\begin{equation}
\label{eq_vertex2dis_asymptotic_naive}
\begin{aligned}
c_{dis,k}(q) \simeq k^{-(2\eta-\bar \eta)}\, u_{dis,k}(\hat q^2)
\end{aligned}
\end{equation}
with the exponent $\bar \eta$ related to the exponent $\theta$ through $\bar \eta= 2+\eta -\theta$ and satisfying $\eta \leq \bar\eta \leq 2\eta$.

In the specific case where the upper bound is saturated, $\bar \eta=2\eta$, the disconnected direct correlation function goes to a constant, just as the matrix-fluid contribution and the two terms must be kept in the expression of $h_{dis,k}(q)$. On the other hand, if $\bar\eta<2\eta$, the matrix-fluid contribution is subdominant when compared to the disconnected fluid-fluid one in the asymptotic regime and can be dropped. This is also true at the higher orders of the HRT hierarchy.

Note that the emergence of a second ``anomalous dimension'' $\bar \eta$ leads to the following behavior of the disconnected pair correlations at criticality:
\begin{equation}
\label{eq_correl_disc_crit}
\begin{aligned}
h_{dis,crit}(q) \sim q^{-(4-\bar \eta)}, 
\end{aligned}
\end{equation}
which implies that the static structure factor of the fluid $S_{ff}(q)=1+\rho h_{con}(q)+ \rho h_{dis}(q)$ also diverges as $q^{-(4-\bar \eta)}$, \textit{i.e.} much more strongly than in the bulk fluid.

To express the HRT hierarchy in the asymptotic regime, it is convenient to introduce ``dimensionless'' quantities. Indeed, it is only by eliminating the scaling dimensions that the scale-free behavior associated with the existence of a fixed point can be found in the RG flow. Near a zero-temperature fixed point, the conventional scaling dimensions are modified\cite{tarjus04} due to the presence of the new exponent $\theta$ (or equivalently $\bar \eta$), and one introduces $\varphi = k^{-(d-4+\bar \eta)/2}(\rho - \rho_{crit})$, $ a_{k}(\varphi) =  k^{-(d-\theta)/2}(\mathcal A_{f,k}(\rho) - \mathcal A_{f,k;crit})$ (where  the subscript $crit$ indicates that a quantity is evaluated at the critical point), etc. All constants, such as $-\beta w_{ff}(q=0)$ in Eq.~(\ref{eq_phi_k_asympt}), can be included in a redefinition of the field as well as of the various functions and vertices; they will no longer appear explicitly.

The asymptotic form of the replica-symmetric HRT hierarchy is derived with the help of above considerations and by dropping all subdominant terms. The first equation then reads 
\begin{equation}
\label{eq_ERGE_A_k_asympt_naive}
\begin{aligned}
&\partial_t a_{k}(\varphi)+ (d-2-\eta + \bar\eta) a_{k}(\varphi) - \frac{1}{2} (d-4+\bar \eta)\varphi \partial_{\varphi }a_{k}(\varphi)\\&= \frac{1}{2} \int_{\hat q} [(2-\eta)r(\hat q^2)- 2 \hat q^2 r'(\hat q^2)]\frac{u_{dis,k}(\hat q^2;\varphi) }{[u_{con,k}(\hat q^2;\varphi)+ r(\hat q^2)]^2}
\end{aligned}
\end{equation}
where $r'(y)$ is the derivative of $r(y)$. The above equation is valid if $\bar\eta<2\eta$. In the case where $\bar\eta=2\eta$, one must also take into account the matrix-fluid contribution, $\rho_0 S_{mm}(0)c_{mf,k}^2$, that is no longer irrelevant compared to $c_{dis,k}$. An illustration of the higher-order equations of the HRT hierarchy is given in appendix~\ref{app:asymptotic_naive}.

\subsection{Replica-symmetric HRT for the RFIM}

The random-field Ising model (RFIM) on a $d$-dimensional lattice is described by the following Hamiltonian\cite{imry-ma75,nattermann98}:
\begin{equation}
\label{eq_hamiltonian_RFIM}
\begin{aligned}
H=-J\sum_{<ij>} S_i S_j -\sum_i h_i S_i,
\end{aligned}
\end{equation}
where $S_i=\pm 1$ are Ising variables, $<ij>$  indicate a sum over distinct pairs of nearest neighbors, and the random field $h_i$ is   delta-correlated in space and is drawn from a symmetric probability distribution with zero mean and variance $\Delta_B$, \textit{i.e.}, $\overline{h_i}=0$, $\overline{h_i h_j}=\delta_{ij} \Delta_B$. Choosing for simplicity a Gaussian distribution and applying the replica trick, one finds that the properties of the model can be obtained from the ``replicated'' Hamiltonian
\begin{equation}
\label{eq_hamiltonian_RFIMreplicated}
\begin{aligned}
H^{rep}=-J\sum_{a=1}^n\sum_{<ij>} S_{ia} S_{ja} -\frac{\Delta_B}{2T}\sum_{a,b=1}^n\sum_i S_{ia} S_{ib},
\end{aligned}
\end{equation}
with, as before,  the $n\rightarrow 0$ limit taken at the end of the calculations\cite{footnote-nongauss}. Note that the replicated model has a global $Z_2$ symmetry, the Hamiltonian being invariant by the simultaneous change $S_{ia}\rightarrow -S_{ia}$ in all replicas. This symmetry is absent in the quenched-annealed mixture at the microscopic level.

One can repeat the steps of the replica-symmetric HRT approach with a change of notations from fluid to magnetic language: the density is replaced by the magnetization $m$=$\overline{<S_{ia}>}$, the Helmholtz free-energy density by the Gibbs free-energy per spin $\mathcal A_{k}(m)$, etc. The reference system is that with only on-site interactions, $-(\Delta_B/2T)\sum_{a,b=1}^n\sum_i S_{ia} S_{ib}$, and the Ising constraint $S_{ia}^2=1$ (which is equivalent in a lattice-gas language to the ``core'' or single-site occupancy condition). The attractive potential is simply $-J \lambda_{ij}$ where $\lambda_{ij}$ is the adjacency matrix of the lattice so that its Fourier transform $\lambda(q)=\sum_{\nu=1}^d \cos(q_{\nu})$ for a $d$-dimensional cubic lattice. An IR regulator is introduced and one defines $\phi_k(q)=R_k(q)\,\beta J  \lambda(q)$.

The HRT hierarchy is then formally the same as that derived above for the quenched-annealed mixture, except that all the direct correlation functions involving the matrix disappear from the flow equations. For instance, the first equation of the HRT hierarchy is simply
\begin{equation}
\label{eq_ERGE_A_k_naiveRFIM}
\begin{aligned}
&\partial_t\mathcal A_{k}(m)=\\&\frac{1}{2} \int_{q} \partial_t \phi_k(q)& \bigg \{ \frac{1}{c_{con,k}(q;m)} + \frac{c_{dis,k}(q;m)}{c_{con,k}(q;m)^2}\bigg\},
\end{aligned}
\end{equation}
where the direct correlation functions are conventionally defined in a slightly different way as in liquid-state statistical mechanics with $c_{con,k}(q;m)=\Gamma_k^{(2)}(q;m)+ \phi_k(q)$ (on the other hand, $c_{dis,k}(q;m)$ has the same definition). With the correspondence $c_{con,k}(q;m) \rightarrow 1/\rho - c_{con,k}(q;\rho)$ [and $c_{dis,k}(q;m) \rightarrow c_{dis,k}(q;\rho)+\rho S_{mm}(q)c_{mf,k}(q;\rho)^2$], Eq.~(\ref{eq_ERGE_A_k_naiveRFIM}) coincides with Eq.~(\ref{eq_ERGE_A_k_naive_RS}). This carries over to the higher orders.

The critical behavior of the RFIM is known to be controlled by a zero-temperature fixed point\cite{villain84,fisher86,middleton-fisher02} and the same scaling as in section~\ref{sub:asymptotic} is  used to put the flow equations in a dimensionless asymptotic form (of course, there is no need here to consider the matrix-fluid functions). These asymptotic flow equations for the RFIM are then identical to those derived above for the quenched-annealed mixture. (They are also identical to the flow equations obtained in the field-theoretical formulation of the RFIM.)

To make fully rigorous the proof that the critical behavior of the quenched-annealed system and that of the RFIM are identical, one must ensure that the initial conditions at the microscopic scale of the two systems do lead to the same fixed point. As previously noted, the symmetry of the initial condition of the RG flow is different in the two cases since the fluid does not have the (statistical) $Z_2$ global inversion symmetry of the Ising model. However, provided that the asymptotic HRT equations (which are formally the same for the RFIM and the quenched-annealed mixture) admit a \textit{unique} fixed point with the appropriate properties to describe a critical point, the flow equations should lose the memory of the initial conditions and the critical behavior of the two systems be the same. (The same is true for the critical behavior of bulk fluids which is in the  same universality class as that of the pure Ising model\cite{HRTreview} despite the absence of $Z_2$ symmetry.) The fact that the (critical) fixed point is unique will be illustrated with specific approximations below. It is also worth pointing out that the above reasoning, as previously discussed, assumes that no accidental cancellation takes place in the quenched-annealed case between disconnected and matrix-fluid components, which should also be checked in actual computations.

\subsection{Ornstein-Zernike approximation}
\label{subsec:naiveOZ/LPA}

Closures of the HRT hierarchy for the quenched-annealed mixture can be devised by extending the approximation schemes developed for the HRT of bulk fluids (see section~\ref{sec:HRTbrief}) to the replicated equilibrium mixture. The simplest  closure is an RPA-like approximation (see section~\ref{sec:HRTbrief}) that enforces the compressibility sum rules, Eqs.~(\ref{eq_compress_connQA_RS}) and (\ref{eq_compress_mfQA_RS}):
\begin{equation}
\label{eq_closure_conn_HRTnaive}
\begin{aligned}
&\frac{1}{\rho}- c_{con,k}(q;\rho)-\phi_{k}(q)=\frac{\partial^2\mathcal A_{f,k}(\rho;\rho_0)}{\partial \rho^2}\\&-[c_{con,R}(q)-c_{con,R}(q=0)]- \left [\phi(q)-\phi(q=0) \right ],
\end{aligned}
\end{equation}
\begin{equation}
\label{eq_closure_mf_HRTnaive}
\begin{aligned}
c_{mf,k}(q;\rho)=-&\frac{\partial^2\mathcal A_{f,k}(\rho;\rho_0)}{\partial \rho \partial \rho_0}\\&+ [c_{mf,R}(q;\rho) -c_{mf,R}(q=0;\rho)],
\end{aligned}
\end{equation}
\begin{equation}
\label{eq_closure_dis_HRTnaive}
\begin{aligned}
c_{dis,k}(q;\rho)=c_{dis,R}(q=0;\rho),
\end{aligned}
\end{equation}
where as before the subscript $R$ indicates a quantity calculated in the reference system in which the fluid atoms interact with the matrix with the full pair potential and interact among themselves through the steep repulsive interaction component.  [Eq.~(\ref{eq_closure_conn_HRTnaive}) for the connected function could also  be replaced by an equation similar to Eq.~(\ref{eq_RPAbis}).]

As for the bulk fluid, the above closure could be improved to better account for the physics at short distances. For instance, if the steep repulsive pair interactions between fluid atoms and between fluid and solid atoms are modeled as hard-core potentials, one can use an ORPA-like closure by adding contributions to the direct correlation functions for distances less than the relevant core diameter: only two core conditions however exist, $h_{ff,k}=h_{con,k}+h_{dis,k}=-1$ inside the fluid-fluid core range and $h_{ff,k}=-1$ inside the fluid-matrix core range. Note also that,  due to the absence of compressibility sum rule relating the disconnected correlation function to the fluid Helmholtz free-energy density, there is no obvious way to improve the closure for this correlation function beyond Eq.~(\ref{eq_closure_dis_HRTnaive}). As a result the disconnected direct correlation function is not renormalized in any of the above approximations.

Irrespective of the degree of refinement, most common approximations of the direct correlation functions are of OZ type, which means that their small-wavevector behavior is regular, \textit{i.e.},
\begin{equation}
\label{eq_OZ_approx}
\begin{aligned}
&\frac{1}{\rho}- c_{con,k}(q;\rho)-\phi_{k}(q)=\frac{\partial^2\mathcal A_{f,k}(\rho;\rho_0)}{\partial \rho^2}+O(q^2),
\\& c_{mf,k}(q;\rho)=-\frac{\partial^2\mathcal A_{f,k}(\rho;\rho_0)}{\partial \rho \partial \rho_0}+O(q^2), 
\\& c_{dis,k}(q;\rho)=c_{dis,R}(q=0;\rho) + O(q^2).
\end{aligned}
\end{equation}
From the regular $q$-dependence of these direct correlation functions, one immediately derives that the critical exponents $\eta$ and $\bar \eta$ satisfy
\begin{equation}
\label{eq_eta_bareta_LPA}
\eta=\bar \eta=0.
\end{equation}

Generically with such an OZ approximation,
\begin{equation}
\label{eq_RSOZ_approx_asympt_con}
\rho^{-1}- c_{con,k}(q;\rho)-\phi_{k}(q)-\partial^2\mathcal A_{f,k}(\rho;\rho_0)/\partial \rho^2\simeq b_k q^2
\end{equation}
and
\begin{equation}
\label{eq_RSOZ_approx_asympt_dis}
c_{dis,k}(q)+\rho_0 S_{mm}(q)c_{mf,k}(q)^2\simeq K_k
\end{equation}
in the small-$k$, small-$q$ limit, where $b_k>0$ and $K_k>0$ are regular functions of $k$, $\rho$, and $T$ (and of the disorder characteristics such as $\rho_0$) with finite limits, $b>0$  and $g>0$ when  $k\rightarrow 0$. (Note that as $\eta=\bar\eta=0$, the matrix-fluid contribution is not subdominant with respect to the disconnected one.)

The asymptotic HRT equation, Eq.~(\ref{eq_ERGE_A_k_asympt_naive}), can be reexpressed as
\begin{equation}
\label{eq_ERGE_A_k_asympt_naiveOZ}
\begin{aligned}
&\partial_t a_{k}(\varphi)+ (d-2) a_{k}(\varphi) - \frac{1}{2} (d-4)\varphi a'_{k}(\varphi)\\&= \frac{1}{2} \int_{\hat q} \frac{2[r(\hat q^2)- \hat q^2 r'(\hat q^2)]}{[\hat q^2+r(\hat q^2)+a_{k}''(\varphi)]^2},
\end{aligned}
\end{equation}
where a prime always denotes a derivative with respect to the argument and the constants $b$ and $K$ have been incorporated into a trivial redefinition of  $a_{k}$ and $\varphi$. It is easy to show that the very same equation as Eq.~(\ref{eq_ERGE_A_k_asympt_naiveOZ}) is obtained under the same assumptions for the RFIM. In the context of field theory, this kind of approximation to the RG flow is known as ``local potential approximation'' (LPA)\cite{wegner,nicoll74}.

For actual computations, we have considered the sharp IR cutoff (see section \ref{sec:HRTbrief}), so that the asymptotic HRT equation becomes
\begin{equation}
\label{eq_ERGE_a_k_naiveOZ_sharp}
\begin{aligned}
\partial_t a_{k}(\varphi)=&- (d-2) a_{k}(\varphi) + \frac{1}{2} (d-4)\varphi a'_{k}(\varphi)\\&+ \left(\frac{1}{1+a''_{k}(\varphi)}-\frac{1}{1+a''_{k}(\varphi=0)} \right ).
\end{aligned}
\end{equation}
The fixed point that describes the gas-liquid critical behavior of the fluid is determined by looking for the scale-free (stationary) solution of the above equation, \textit{i.e.} $\partial_t a_{k}\vert_*=0$. Taking for convenience one derivative with respect to the field, we then obtain
\begin{equation}
\label{eq_ERGE_a'FP_LPAsharp_naive}
\begin{aligned}
d a'_{*}(\varphi) - (d-4)\varphi a''_{*}(\varphi) + 2\frac{a'''_{*}(\varphi)}{[1+a''_{*}(\varphi)]^2}=0.
\end{aligned}
\end{equation}
One has to look for a solution that behaves at large positive and negative field ($\varphi \rightarrow \pm \infty$) as $a'_{*}(\varphi)\sim \vert \varphi \vert^{d/(d-4)}$ and that is defined and continuous over the whole real axis. This condition ensures that the second-order differential equation has only a small number of acceptable solutions\cite{morris94}.

When the spatial dimension $d\geq 6$, it is easily shown that the only fixed-point solution is $a_*(\varphi)=0$, which corresponds to the Gaussian fixed point. The exponents are then the classical ones obtained from mean-field theory; $d=6$ is therefore the upper critical dimension of the system at criticality, which represents a shift by $2$ compared to bulk fluids and results from the strong effect of the quenched disorder (random chemical potential or random field).  Below $6$, a new nontrivial solution with properties associated to a critical point is found in addition to the Gaussian fixed point, and the latter becomes unstable.

The fact that the anomalous dimensions $\eta$ and $\bar\eta$ are zero in the present approximation prevents us from studying dimensions below $4$: as can be seen from Eq.~(\ref{eq_correl_disc_crit}), which is equivalent in real space to $h_{dis,crit}(r) \sim r^{-(d-4+\bar \eta)}$, the critical decay of the pair correlations is not ensured in this case and, in fact, no critical point is expected. We have therefore calculated the solutions of Eq.~(\ref{eq_ERGE_a'FP_LPAsharp_naive}) in $d=5$ by using a shooting method as in ref.~[\onlinecite{morris94}]. We have found a unique nontrivial solution $a'_*(\varphi)$ satisfying the conditions described above, and it is an odd function of the field. The corresponding $a_*(\varphi)$ is then an even function. As for the bulk fluid whose critical behavior is in the same universality class as the Ising model, one recovers that the $Z_2$ inversion symmetry (which for liquids amounts to a symmetry with respect to the critical density) is obeyed by the critical theory, as in the RFIM, despite the fact that the initial Hamiltonian of the quenched-annealed model is not symmetric.

The stability of the fixed point and the critical exponents (if the fixed point indeed properly describes the critical behavior) are investigated by linearizing  the flow equation around the fixed-point solution and studying the spectrum of eigenvalues. Because of the asymptotic $Z_2$ symmetry of the fixed point, the eigenvalues can be sorted according to the odd or even behavior under field $\varphi$ reversal of the associated eigenfunctions. The largest eigenvalue $\Lambda_e$ corresponding to an even eigenfunction describes the relevant direction associated with the temperature or the disorder strength whereas the largest eigenvalue $\Lambda_o$ associated with an odd eigenfunction describes the chemical-potential (or magnetic-field for the RFIM) dependence along the critical isotherm. All other eigenvalues should be negative for a proper critical fixed point. Writing $a_k(\varphi)=a_*(\varphi) + k^{-\Lambda}f_{\Lambda}(\varphi)$ and then linearizing around the fixed point lead to
\begin{equation}
\label{eq_ERGE_eigenvalue_LPAsharp_naive}
\begin{aligned}
&(-\Lambda +d-2)f_{\Lambda}(\varphi) - \frac{d-4}{2}\varphi f'_{\Lambda}(\varphi) + \frac{f''_{\Lambda}(\varphi)}{[1+a''_{*}(\varphi)]^2}\\&=\frac{f_{\Lambda}(0)}{[1+a''_{*}(0)]^2}.
\end{aligned}
\end{equation}
The ``odd'' eigenvalue $\Lambda_o$ is obtained analytically by noticing that the above equation has an odd solution $f_{\Lambda}(\varphi)=a'_*(\varphi)$ with $\Lambda_o=(d-4)/2$ [compare with Eq.~(\ref{eq_ERGE_a'FP_LPAsharp_naive})]. The ``even'' eigenvalue $\Lambda_e$ and its associated eigenfunction on the other hand must be numerically determined. (One looks for an even solution of the second-order differential equation that is, as before, defined and continuous everywhere but behaves now as $\varphi \vert^{2(d-2+\Lambda)/(d-4)}$ when $\varphi \rightarrow \infty$.) One then obtains from these eigenvalues the following critical exponents in $d=5$:
\begin{equation}
\label{eq_exponents_LPAsharp_naive}
\begin{aligned}
&\nu=\frac{1}{\Lambda_e}\simeq 0.6496,\\&
\gamma=2\nu\simeq 1.299,\\&
\delta=1+\frac{2}{\Lambda_o}=\frac{d}{d-4}=5,
\end{aligned}
\end{equation}
where $\gamma$ is the exponent describing the divergence of the connected susceptibility $(\partial^2 \mathcal A_f/\partial \rho^2)^{-1}$, whereas  $\delta$ characterizes the relation between density and chemical potential along the critical isotherm.  (Recall that $\eta=\bar\eta=0$ so that $\theta=2$; one can also define an exponent for the divergence of the ``disconnected susceptibility'', $\bar \gamma=(4-\bar\eta)\nu\simeq 2.598$.) In addition, we have computed the second largest eigenvalue corresponding to an even eigenfunction. It is negative, as required, and its absolute value provides the critical exponent $\omega$ controlling the corrections to scaling. The result is
\begin{equation}
\label{eq_exponent_omega_LPAsharp_naive}
\begin{aligned}
\omega \simeq 0.6557.
\end{aligned}
\end{equation}

It is puzzling to find that the exponents predicted by the above OZ approximation for the critical behavior of a fluid in a disordered porous matrix (and equivalently the LPA for the RFIM) do not obey the so-called ``dimensional reduction''\cite{nattermann98}. According to the latter property, the critical behavior of the RFIM in dimension $d$ is the same as the critical behavior of the pure Ising model without quenched disorder in dimension $d-2$. The critical exponents for the system in the absence of quenched disorder (bulk fluid or pure Ising model) in $d= 3$ are equal to
\begin{equation}
\label{eq_exponents_LPAsharp_bulk}
\begin{aligned}
\nu\simeq 0.689, \gamma \simeq1.378, \delta=5, \omega \simeq 0.581
\end{aligned}
\end{equation}
in the OZ/LPA approximation with a sharp IR cutoff\cite{parola-reatto85,HRTreview}. Except for $\delta$, they are thus different from the exponents obtained above for $d=5$ (even when considering error bars).

The dimensional-reduction property of the RFIM is predicted by perturbation theory at all orders\cite{aharony76,grinstein76,young77,parisi79}, but has been rigorously proven wrong in low enough dimensions (\textit{e.g.}, in $d=3$)\cite{imbrie84,bricmont87}. However, from previous studies on the RFIM, one does \textit{not} expect that the breakdown of dimensional reduction could be captured by the simple replica-symmetric formulation used above\cite{RSBmezard,RSBdedom}. One indeed finds that dimensional reduction is verified at first order in the expansion in $\epsilon=6-d$ around the upper critical dimension in the OZ/LPA approximation: in this case, one easily obtain from Eq.~(\ref{eq_ERGE_a'FP_LPAsharp_naive}) that $a_*(\varphi)\propto \epsilon(\varphi^2/2-\varphi^4/12)$; linearizing around this fixed point leads to
\begin{equation}
\label{eq_exponents_LPAsharp_epsilon}
\begin{aligned}
\nu=\frac{1}{2}+\frac{\epsilon}{12},\, \gamma=1+\frac{\epsilon}{6},\, \delta=3+\epsilon,
\end{aligned}
\end{equation}
which exactly corresponds to the result at first order in the expansion in $\epsilon=4-d$ around the upper critical dimension of the pure Ising model or the bulk fluid\cite{HRTreview}. At second order in the $\epsilon$-expansion however, the dimensional-reduction property is lost in the present approximation\cite{footnote-dimred} while it is still verified in the exact perturbative expansion. Clearly the breakdown of dimensional reduction in the present approach is a consequence of the approximations made, not a genuine prediction.

\section{Flaws of the naive HRT  and lessons from the RFIM}
\label{sec:flaws}

\subsection{Flaws of the naive HRT approach}

The naive HRT approach presented above has three major flaws:

(1) The fact that dimensional reduction is not obeyed, despite the simple nature of the approximation that cannot describe the complex physics associated with dimensional-reduction breakdown (metastable states for instance\cite{parisi84b}), is likely to be an artifact of the chosen approximation. In the case of the RFIM, dimensional reduction has been shown to follow from an underlying ``supersymmetry'' of the field theory describing the long wavelength behavior at criticality\cite{parisi79}. As we have obtained indications that the critical behavior of a fluid in a disordered porous material and that of the RFIM are the same, we can anticipate that the same underlying supersymmetric property could in principle be asymptotically found in the quenched-annealed mixture. Clearly, one should then distinguish between a \textit{spontaneous} breaking of the supersymmetry, which is the physical mechanism for the (necessary) breakdown of dimensional reduction in low dimensions (at least in $d=2,3$), and an \textit{explicit} breaking of supersymmetry due to an inappropriate choice of approximation and cutoff functions. The naive approach most likely breaks the supersymmetry explicitly. However there is no way to detect, and further to cure, this problem within the formalism so far developed.

(2) The OZ/LPA approximation does not allow for field renormalization, which implies that $\eta=\bar \eta=0$. As the spatial dependence of the $2$-point Green's function/total pair correlation function at the critical point are predicted to go as $r^{-(d-4+\bar \eta)}$ for the disconnected component, no criticality is found for dimensions $d<4$ and one cannot directly study the physically relevant three-dimensional situation. It is known that going beyond the OZ/LPA approximation with a hard-cutoff formulation is difficult due to the proliferation of singularities associated with derivatives of the cutoff function.

(3) The replica-symmetric formalism, besides being wrong when a hypothetical spontaneous breaking of the permutational symmetry between replicas takes place, does not allow one to easily close the equation at the level of the disconnected $2$-point vertices. As we have seen, there is no compressibility sum rule relating the latter to the average Helmholtz free-energy functional of the fluid. As a result, we had to assume that the disconnected fluid-fluid direct pair correlation function is not renormalized and essentially remains equal to its values in the reference system. (On the other hand, the matrix-fluid direct pair correlation function can be renormalized but the effect of this renormalization is benign as $\partial^2 \mathcal A_f/\partial \rho \partial \rho_0$ stays finite at criticality.)

Going beyond the naive HRT thus requires to find ways for addressing these  shortcomings.

\subsection{A detour via the RFIM and superfield theory}
\label{sec:superfield}

It actually took a long detour via the (super)field theory of the RFIM and its nonperturbative RG description\cite{tarjus04,tissier06,tissier11} to make progress. To keep a long story short we only briefly summarize the lessons that can be drawn from the latter study.

First, the replica-symmetric formalism should be replaced by a framework in which the replica symmetry is explicitly broken by letting the sources (in field-theoretical language) or chemical potentials (in fluid language) be different for each replica. This \textit{a priori} unphysical situation (the fluid in the disordered porous medium or the Ising model in a random field are subject to a unique chemical potential or external magnetic field) is actually a means to properly generate the ``cumulants of the renormalized disorder''. Indeed, in the presence of quenched disorder (and before any use of a replica trick), the generating functionals and thermodynamic potentials of the fluid (or Ising model) are random quantities. As such, they should be described by a functional probability distribution or by an infinite set of cumulants, the latter being more convenient as the cumulants are averaged quantities for which statistical invariance under translations and rotations is recovered. It is easily realized that the only way to describe the cumulants of order $2$ or more with their full functional dependence is by considering copies or replicas of the initial system with the \textit{same} disorder but different density fields, hence different chemical potentials (or sources). This point is central as it has been found in the case of the RFIM that the influence of the rare collective events stemming from the presence of quenched disorder (abrupt changes of the ground-state configuration under a variation of the external source known as  ``avalanches'' or ``shocks'' and near-degenerate metastable states known as ``droplets'') precisely results in a singular behavior of the second and higher-order cumulants in their functional dependence\cite{tarjus04,tissier06,tissier11}. This will be made more explicit below.

Secondly, the underlying supersymmetry does not show up easily in the conventional field formalism, even when one considers an explicit breaking of the replica symmetry. It is made explicit when one upgrades the description to a superfield approach that arises for formal manipulations of the original problem\cite{parisi79}. There is no point to go here through the derivation. It suffices to say that the  supersymmetry can be viewed as an asymptotic property of the theory and that its consequences in the present formalism are exact relations between 1PI vertices (direct correlation functions) that are known as Ward-Takahashi identities\cite{zinnjustin89}. For instance supersymmetry in the RFIM implies that asymptotically, \textit{i.e.} near the critical point and at small momenta (long wavelengths)\cite{tissier11},
\begin{equation}
\label{eq_ward_susy_momentum_2}
\Gamma_{k2}^{(11)}(q^2;m,m) = \Delta\, \partial_{q^2} \Gamma_{k1}^{(2)}(q^2;m),
\end{equation}
with the obvious notation: $\Gamma_{k2;q_1,q_2}^{(11)}(m,m)=(2\pi)^d \delta^{(d)}(q_1+q_2)\Gamma_{k2}^{(11)}(q_1^2;m,m)$, etc. In the above equation, the magnetization $m$ is taken as uniform (and the system is then uniform and isotropic at large distance). The constant $\Delta$ is the variance of the random field, possibly renormalized to include fluctuations from the microscopic scale down to the beginning of the asymptotic regime. Identities similar to Eq.~(\ref{eq_ward_susy_momentum_2}) are also derived for the higher orders.

For the fluid in the presence of a disordered porous material, Ward-Takahashi identities analogous to those for the RFIM should apply in the asymptotic regime, provided of course the supersymmetry is not spontaneously broken\cite{tissier11}. For instance, when $k\rightarrow 0$ and $q\rightarrow 0$ (and if $\bar \eta<2 \eta$), one expects that, similarly to Eq.~(\ref{eq_ward_susy_momentum_2}), supersymmetry implies that the disconnected direct pair correlation function (whose definition in the formalism with explicit replica-symmetry breaking will be made precise below) satisfies
\begin{equation}
\label{eq_ward_susy_momentum_2QA}
c_{dis,k}(q^2;\rho,\rho;\rho_0) \simeq K \, \partial_{q^2}c_{con,k}(q^2;\rho;\rho_0),
\end{equation}
with $K$ a constant $>0$ which is essentially the quantity $c_{dis,k}(q)-\rho_0 S_{mm}(q)c_{mf,k}(q)^2$ evaluated for $q=0$ and renormalized only down to some $k$ at the beginning of the asymptotic regime. 

Finally, in order to beyond the OZ/LPA approximation and introduce a renormalization of the field that allows us to study criticality in three dimensions, it is more convenient to replace the hard IR cutoff by a smooth one\cite{berges02}. Approximation schemes then exist for improving the description of  the long-distance spatial dependence of the correlation functions and provide accurate values of the anomalous dimensions $\eta$ and $\bar \eta$\cite{tarjus04,tissier06,tissier11}.
 
\section{Toward a proper HRT for fluids in disordered media}
\label{sec:properHRT}

\subsection{Explicit replica-symmetry breaking}
\label{sub:explicitRSB}

To adapt the formalism with an explicit breaking of the replica symmetry (see above)\cite{tarjus04}, we start with the grand partition function(al) of the fluid in the presence of a given configuration of the solid matrix characterized by the microscopic density $\hat \rho_0$:
\begin{equation}
\label{eq_Xi_RSB}
\begin{aligned}
&\Xi[\mu;\hat \rho_0]=\exp \left (W[\mu;\hat \rho_0]\right ) \\& = \sum_{N} \frac{1}{N!} \int_{x_{1}}...\int_{x_{N}} \exp \big\{-\beta  V_R(\{x_{i}\}_N;\hat \rho_0) +\\&  \frac{1}{2}  \int_{x} \int_{y} \phi(\vert x - y\vert)\hat \rho(x) \hat \rho(y)+\beta   \int_{x} \mu(x) \hat \rho(x) \big\}
\end{aligned}
\end{equation}
where as before $V_R(\{x_{ai}\}_N;\hat \rho_0)$ contains the sum of all pairwise steep repulsive fluid-fluid interaction $v_R$ among atoms of replica $a$ as well as the sum of all pairwise interactions $v_{mf}$ between the fluid replica $a$ and the solid matrix, and  $\phi=-\beta w_{ff}$.

We next consider $n$ copies or replicas of the fluid in the presence of the \textit{same} matrix configuration but coupled to \textit{different} chemical potentials and we define
\begin{equation}
\label{eq_W_RSB}
\begin{aligned}
&\exp W[\{\mu_a\};\rho_0 ] = \overline{ \prod_{a=1}^{n} \Xi[\mu_a;\hat \rho_0]}=\overline{\exp\left (\sum_{a=1}^{n} W[\mu_a;\hat \rho_0]\right )}
\end{aligned}
\end{equation}
where the overline denotes an average over the disorder, \textit{i.e.} the configurations of matrix particles $\hat \rho_0$; $\rho_0=\overline{\hat \rho_0}$ is the average solid density and $\mu_a$ is the chemical potential for replica $a$. The functional $W[\{\mu_a\};\rho_0 ]$ can be expanded in cumulants of the random functional $W[\mu;\hat \rho_0]$:
\begin{equation}
\label{eq_W_cumulants}
\begin{aligned}
&W[\{\mu_a\};\rho_0 ] =\\&
\sum_{a=1}^{n} \overline{W[\mu_a;\hat \rho_0]}+ \frac{1}{2} \sum_{a,b=1}^{n}\overline{W[\mu_a;\hat \rho_0]W[\mu_b;\hat \rho_0]}\big \vert_{cum}+\\& \frac{1}{3!} \sum_{a,b,c=1}^{n}\overline{W[\mu_a;\hat \rho_0]W[\mu_b;\hat \rho_0]W[\mu_c;\hat \rho_0]}\big \vert_{cum}+ \cdots
\end{aligned}
\end{equation}

A convenient trick to extract the cumulants of $W[\mu_a;\hat \rho_0]$ with their full functional dependence (\textit{i.e.}, generically, with different arguments $\mu_a$, $\mu_b$, etc.) is then to let the number of replicas be arbitrary and to view the expansion in the right-hand side of Eq.~(\ref{eq_W_cumulants}) as an expansion in increasing number of unconstrained, or ``free'', sums over replicas\cite{tarjus04,largeNledoussal03,largeNledoussal04}:
\begin{equation}
\label{eq_W_free_replica_sums}
\begin{aligned}
W[\{\mu_a\};\rho_0 ] =&
\sum_{a=1}^{n}W_1[\mu_a;\rho_0 ]+ \frac{1}{2} \sum_{a,b=1}^{n}W_2[\mu_a,\mu_b;\rho_0 ]\\& +\frac{1}{3!} \sum_{a,b,c=1}^{n}W_3[\mu_a,\mu_b,\mu_c;\rho_0 ]+ \cdots,
\end{aligned}
\end{equation}
where the term of order $p$ in the expansion is a sum over $p$ replica indices of a functional $W_p$ depending exactly on $p$ replica sources, this functional being precisely equal here to the $p$th cumulant of $W[ \mu; \hat \rho_0]$. This procedure, which rests on an explicit breaking of the replica symmetry through the introduction of the chemical potentials $\mu_a$ is \textit{a priori} different from the standard use of replicas, in which all chemical potentials (or sources in field-theoretical language) are equal;  it therefore avoids the delicate handling of a spontaneous replica symmetry breaking when present\cite{mouhanna10}. The first cumulant $W_1$ gives the grand potential of the fluid, $W_1[\mu;\rho_0 ]=-\beta \Omega_f[\mu;\rho_0 ]$ and the higher-order cumulants provide information on the probability distribution of the random functional $W[\mu;\hat \rho_0]$.

An effective action or Helmholtz free-energy functional is as usual defined through a Legendre transform,
\begin{equation}
\label{eq_Gamma_QA_RSB}
\begin{aligned}
&\Gamma[\{\rho_a\};\rho_0 ]= -W[\{\mu_a\};\rho_0 ]+ \beta \sum_{a=1}^{n}  \int_{x} \mu_a(x) \rho_a(x).
\end{aligned}
\end{equation}
where 
\begin{equation}
\label{eq_rho_legendre_RSB}
\begin{aligned}
\rho_a(x)= \frac{\delta W[\{\mu_a\};\rho_0 ]}{\beta \delta \mu_a(x)}
\end{aligned}
\end{equation}
is the fluid-density field in replica $a$. (As before, if necessary, one can take into account self-interaction terms by replacing 
$\mu_a(x)$ by $\mu_a(x) -w_{ff}(0)/2$.)

Just as the functional $W$, the effective action can be expanded in increasing number of free replica sums,
\begin{equation}
\begin{split} 
 \label{eq_Gamma_cumulants}
\Gamma [\left\lbrace \rho_a\right\rbrace;\rho_0 ]= & \Gamma_0[\rho_0 ]+\sum_{a=1} ^{n} \Gamma_1[ \rho_a;\rho_0 ] -\dfrac{1}{2}\sum_{a,b=1} ^{n}\Gamma_2[ \rho_a,  \rho_b;\rho_0 ]\\& + \dfrac{1}{3!}\sum_{a,b,c=1} ^{n}\Gamma_3[ \rho_a,  \rho_b,  \rho_c;\rho_0 ] + \cdots,
\end{split}
\end{equation}
where for later convenience we have introduced a minus sign for all even terms of the expansion. We have added a first term, $\Gamma_0$, that only depends on the solid matrix and is useful to generate the direct correlation functions (1PI vertices) of the matrix\cite{footnote-matrix}.

$\Gamma [\left\lbrace  \rho_a\right\rbrace;\rho_0  ]$ and $W[\left\lbrace  \mu_a\right\rbrace;\rho_0  ]$ are related by a Legendre transform, so if one also expands the sources $ \mu_a[\left\lbrace  \rho_f\right\rbrace;\rho_0  ]$ (where we have denoted $\left\lbrace  \rho_f\right\rbrace$ the $n$ replica fields to avoid confusion in the indices) in increasing number of free replica sums, one can relate the terms of the expansion of the effective action to the cumulants of the random functional $W[ J; h]$. The relation is straightforward for the first terms, but gets more involved as the order increases\cite{tarjus04}.

More precisely, $\Gamma_1[ \rho;\rho_0 ]$ is the Legendre transform of $W_1[ J;\rho_0 ]$, namely,
\begin{equation}
\label{legendre_gamma_1}
\Gamma_1[ \rho;\rho_0  ] = - W_1[ \mu;\rho_0 ] +  \int_{ x}  \mu( x) \rho( x),
\end{equation}
with
\begin{equation}
\label{legendre_rho}
\rho( x)=\frac{\delta W_1 [ \mu ;\rho_0 ] }{\delta  \mu( x)},
\end{equation}
and the second-order terms is given by
\begin{equation}
  \label{eq_cumg2}
\Gamma_2[ \rho_1,  \rho_2;\rho_0 ] = W_2[ \mu[ \rho_1;\rho_0 ],  \mu[ \rho_2;\rho_0 ];\rho_0 ],
\end{equation}
where  $ \mu[ \rho;\rho_0 ]$ is the \textit{nonrandom} chemical potential defined via the inverse of the Legendre transform relation in Eq.~(\ref{legendre_gamma_1}), \textit{i.e.},
\begin{equation}
\label{eq_legendre_rho_mu}
\mu[ \rho;\rho_0]( x)=\frac{\delta \Gamma_1 [  \rho;\rho_0]} {\delta \rho( x)}.
\end{equation}
(Note that $ \mu[\rho_a;\rho_0](x)$ introduced here differs from the source $ \mu_a(x)$ appearing in Eq.~(\ref{eq_Gamma_QA_RSB}): through the Legendre relations, the latter depends on all the replica fields $\{\rho_f\}$ while the former depends on a single replica field $\rho_a$.) The above expression motivates our choice of signs for the terms of the expansion in free replica sums of $\Gamma [\left\lbrace  \rho_a\right\rbrace;\rho_0  ]$, Eq.~(\ref{eq_Gamma_cumulants}): $\Gamma_2[ \rho_1,  \rho_2;\rho_0 ]$ is directly the second cumulant of $W[ \mu; \hat\rho_0]$ (with the proper choice of $ \mu[ \rho;\rho_0 ]$). 

We point out that $\Gamma_p[ \rho_1, ...,  \rho_p]$ for $p\geq 3$ cannot be directly taken as the $p$th  cumulant of a physically accessible random functional, in particular not of the disorder-dependent Legendre transform of $W[\mu; h]$ (although it can certainly be expressed in terms of such cumulants of order equal to, or lower than, $p$). In the following and by an abuse of language, we will nonetheless casually refer to the $\Gamma_p$'s as ``cumulants of the renormalized disorder'' (which is true for $p=2$).

\subsection{HRT with an explicit breaking of replica symmetry}

We are now in a position to formulate a more powerful HRT framework for the quenched-annealed mixture. This first goes through the definition of a grand partition function(al) at the scale $k$,
\begin{equation}
\label{eq_Xi_k_RSB}
\begin{aligned}
&\Xi_k [\{\mu_a\};\hat \rho_0]=\exp \left (W_k[\{\mu_a\};\hat \rho_0]\right ) = \sum_{N_1,...,N_n}\frac{1}{N_1!...N_n!}\\&\int_{\{x_{1i}\}_{N_1}}...\int_{\{x_{ni}\}_{N_n}} \exp \big\{-\beta \sum_{a=1}^{n} V_R(\{x_{ai}\}_{N_a};\hat \rho_0) + \\&  \frac{1}{2} \sum_{a,b=1}^{n} \int_{x}\int_{y}\big [\phi(\vert x - y\vert)\delta_{ab} - \phi_{k,ab}(\vert x - y\vert)\big]\hat \rho_a(x) \hat \rho_b(y) \\&+\beta  \sum_{a=1}^{n} \int_{x} \mu_a(x) \hat \rho_a(x) \big\},
\end{aligned}
\end{equation}
where $\phi_{k,ab}(q)= - \beta w_{k,ab}$ with $w_{k,ab}(q)=\mathcal R_{k,ab}(q) w_{ff}(q)$, $a,b=1,...,n$. Note that we do not consider any cutoff function for the matrix-matrix and  matrix-fluid interactions.

We next introduce the generating functional of the Green's functions at scale $k$, $W_k[\{\mu_a\}; \rho_0]$, through
\begin{equation}
\label{eq_W_k_RSB}
\begin{aligned}
&\exp W_k[\{\mu_a\}; \rho_0] = \overline{\Xi_k [\{\mu_a\};\hat \rho_0]}=\overline{\exp\left (W_k[\{\mu_a\};\hat \rho_0]\right )}
\end{aligned}
\end{equation}
and the effective average action (Helmholtz free-energy functional) at scale $k$, $\Gamma_k[\{\rho_a \};\rho_0]$, through the modified Legendre transform
\begin{equation}
\label{eq_Gamma_k_QA_RSB}
\begin{aligned}
&\Gamma_k[\{\rho_a\}; \rho_0]= -W_k[\{\mu_a\}; \rho_0]+ \beta \sum_{a=1}^{n}  \int_{x} \mu_a(x) \rho_a(x)\\&+\frac{1}{2} \sum_{a,b=1}^{n} \int_{x}\int_{y} \phi_{k,ab}(\vert x - y\vert)\rho_a(x)\rho_b(y).
\end{aligned}
\end{equation}

The evolution with $k$ of the effective average action again follows an exact flow equation,
\begin{equation}
\label{eq_ERGE_QA}
\begin{aligned}
&\partial_t\Gamma_k[\{\rho_a \};\rho_0]=\\&\frac{1}{2} \sum_{a,b=1}^{n}\int_{x}\int_{y} \partial_t \phi_{k,ab}(\vert x-y\vert)  F_{k,ab}[x,y;\{\rho_a\};\rho_0],
\end{aligned}
\end{equation}
where the $2$-point (Green's) correlation matrix at scale $k$, $ \vect F_k=\vect W_k^{(2)}$, is the inverse in the sense of matrices and operators of $\Gamma_{k;ab}^{(2)}[x,y;\{ \rho_a \};\rho_0]+ \phi_{k;ab}(\vert x-y\vert)$.

The reasoning developed in the previous subsection can be applied to the effective average action $\Gamma_k$ and its expansion in free replica sums. As a results, Eqs.~(\ref{eq_W_cumulants}) to (\ref{eq_legendre_rho_mu}) can be extended to any running scale $k$. To make the expansion in free replica sums a fully operational procedure, one needs to be able to perform systematic algebraic manipulations, as for instance the inversion of the matrix $ \vect \Gamma_k^{(2)}+\vect \phi_k$. We detail in appendix~\ref{app:algebraic_replica_sums} the method for matrices that depend on two replica indices but are functionals of the $n$ replica fields; extension to higher-order tensors is presented in Ref.~[\onlinecite{largeNledoussal04}].

From the detour via the field theory of the RFIM and its supersymmetric formulation\cite{tissier11}, we know that, contrary to what was done in the naive HRT approach,  the matrix $\phi_{k,ab}$ cannot be simply taken as diagonal despite the fact that there is no direct interaction between atoms belonging to different replicas. The replicas are indeed indirectly correlated through the interaction with the quenched matrix and the corresponding fluctuations must also be regularized. We have seen in section~\ref{sec:superfield} that this is necessary to avoid an explicit breaking of the underlying supersymmetry. We therefore choose $\phi_{k,ab}(q)= - \beta  R_{k,ab}(q) w_{ff}(q)$ with
\begin{equation}
\label{eq_regulator_QA}
\begin{aligned}
 R_{k,ab}(q)=\widehat R_k(q) \delta_{ab} + \widetilde R_k(q)
\end{aligned}
\end{equation}
with $\widetilde R_k(q) \propto - \partial_{q^2}\widehat R_k(q)$. We shall denote by $\widehat \phi_k$ and $\widetilde \phi_k$ the corresponding elements of the matrix $\phi_k$.

Using the results of the previous section and of appendix~\ref{app:algebraic_replica_sums}, we derive through systematic expansions in free sums over replicas a hierarchy of exact HRT flow equations for the ``cumulants'' associated with the effective average action (Helmholtz free-energy functional). The functional equation for the first cumulant reads
\begin{equation}
\label{eq_flow_Gamma1}
\begin{aligned}
\partial_t\Gamma_{k1}\left[ \rho_1;\rho_0 \right ]=
\dfrac{1}{2} \int_{ q} & \bigg \{ \partial_t (\widehat{\phi}_k(q)+ \widetilde{\phi}_k(q))  F_{con,k}\left[ \rho_1;\rho_0 \right]_ {-q\,  q} \\& +\partial_t \widehat{\phi}_k(q) F_{dis,k}\left[ \rho_1, \rho_1 ;\rho_0 \right ]_ {-q\,  q}\bigg \} ,
\end{aligned}
\end{equation}
where the correlation functions (or Green's functions or else ``propagators'') $F_{con,k}\equiv \widehat F_k$ and $F_{dis,k}\equiv \widetilde F_k$ are obtained as zeroth-order components of the expansion in number of free replica sums of the matrix $\vect F_k$; they are explicitly given by
\begin{equation}
\label{eq_hatpropagator}
F_{con,k}[  \rho_1;\rho_0  ]=\left(  \Gamma _{k1}^{(2)}[  \rho_1;\rho_0  ]+\widehat \phi_k  \right) ^{-1},
\end{equation}
\begin{equation}
\begin{aligned}
\label{eq_tildepropagator}
&F_{dis,k}[  \rho_1, \rho_2;\rho_0 ]= F_{con,k}[  \rho_1 ]\bigg( \Gamma _{k2}^{(11)}[  \rho_1, \rho_2;\rho_0  ]+\rho_0  \times \\&\Gamma _{k1}^{(1;1)}[  \rho_1;\rho_0 ]S_{mm}[\rho_0 ]\Gamma _{k1}^{(1;1)}[  \rho_2;\rho_0 ]-\widetilde \phi_k \bigg) F_{con,k}[  \rho_2;\rho_0  ],
\end{aligned}
\end{equation}
where all quantities have to be considered as operators in real or Fourier space when the density fields are inhomogeneous. The above expressions follow directly from the results derived in appendix~\ref{app:algebraic_replica_sums}. Note that the $2$-point 1PI vertex $\Gamma _{k1}^{(1;1)}[\rho_1;\rho_0]$ only involves the first, $1$-replica, cumulant and corresponds to a matrix-fluid direct correlation function in the language used in sections~\ref{sec:QAreplicas} and \ref{sec:naive}.

The functional HRT equation describing the flow of the second cumulant can be expressed as
\begin{equation}
\label{eq_flow_Gamma2_final}
\begin{split}
&\partial_t \Gamma_{k2}\left[ \rho_1 , \rho_2;\rho_0\right ]= \dfrac{1}{2} \widetilde{\partial}_t  Tr \bigg \{\widehat{F}_{k}\left[ \rho_1;\rho_0 \right ] \big( \Gamma _{k2}^{(20)}\left[ \rho_1,  \rho_2;\rho_0 \right ] -\\&  -  \Gamma _{k3}^{(110)}\left[ \rho_1,  \rho_1,  \rho_2 ;\rho_0\right ]\big ) +\widetilde{ F}_{k}\left[ \rho_1, \rho_1;\rho_0 \right ]  \Gamma _{k2}^{(20)}\left[ \rho_1, \rho_2;\rho_0 \right ]  +\\&\dfrac{1}{2} \widetilde{ F}_{k}\left[ \rho_1, \rho_2;\rho_0 \right ] ( \Gamma _{k2}^{(11)}\left[ \rho_1, \rho_2;\rho_0 \right ]+\rho_0 \Gamma _{k1}^{(1;1)}[  \rho_1;\rho_0 ]S_{mm}[\rho_0 ]\times \\& \Gamma _{k1}^{(1;1)}[  \rho_2;\rho_0 ] - \widetilde{\phi}_k)  +\dfrac{\rho_0}{2}\widetilde{ F}_{k}\left[ \rho_1, \rho_2;\rho_0 \right ] \Gamma _{k1}^{(1;1)}[  \rho_1;\rho_0 ]S_{mm}[\rho_0] \\& \times \big (\Gamma _{k2}^{(10;1)}[  \rho_1,\rho_2;\rho_0 ]+ \dfrac{\rho_0}{2} \Gamma _{k1}^{(0;1)}[  \rho_2;\rho_0 ]S_{mm}[\rho_0] \Gamma _{k1}^{(1;1)}[  \rho_2;\rho_0 ]\big)\\&+ perm (12) \bigg \},
\end{split}
\end{equation}
where $perm (12)$ denotes the expression obtained by permuting $ \rho_1$ and $ \rho_2$ (some care is needed in the term by term identification in order to properly symmetrize the expressions and satisfy the permutational property of the various arguments of the cumulants), the operator $\widetilde{\partial}_t$ acts only on the cutoff functions (with \textit{e.g.}, $\widetilde{\partial}_t \widehat{F}_{k;qq'}=-\int_{q''}\widehat{F}_{k;q''}\partial_t \widehat{R}_{k}(q'')\widehat{F}_{k;q''q'}$), and the trace $Tr$ is over momenta. Similar equations can be derived for the higher-order cumulants. 

As in the case of the naive HRT, the hierarchy for the cumulants of the renormalized disorder, when expressed in a functional form as above, contains full information on the complete set of 1PI (or direct) correlation functions. Flow equations for the latter are simply obtained by taking appropriate functional derivatives. 

We conclude this subsection by providing the counterpart of Eqs~(\ref{eq_hatpropagator}) and (\ref{eq_tildepropagator}), \textit{i.e.}, the OZ equations for the quenched-annealed mixture in the explicit replica-symmetry breaking framework. We consider for simplicity uniform replica density fields, which leads to
\begin{equation}
\label{eq_OZ_RSB_con}
1 + \rho_1 h_{con,k}(q;\rho_1;\rho_0)=\frac{1}{1-\rho_1 c_{con,k}(q;\rho_1;\rho_0)},
\end{equation}
\begin{equation}
\begin{aligned}
\label{eq_OZ_RSB_dis}
&h_{dis,k}( q;\rho_1, \rho_2;\rho_0)=\big [1+\rho_1 h_{con,k}(q;\rho_1;\rho_0)\big ]\times \\& \big [1+\rho_2 h_{con,k}(q;\rho_2;\rho_0)\big ] \big[c_{dis,k}( q;\rho_1, \rho_2;\rho_0)+\rho_0 S_{mm}(q;\rho_0)\\& \times c _{mf,k}(q;\rho_1;\rho_0)c _{mf,k}(q;\rho_2;\rho_0)\big ],
\end{aligned}
\end{equation}
whereas, the matrix fluid total pair correlation function is given by
\begin{equation}
\begin{aligned}
\label{eq_OZ_RSB_mf}
h_{mf,k}( q;\rho_1;\rho_0)=S_{mm}(q;\rho_0)\frac{c _{mf,k}(q;\rho_1;\rho_0)}{1-\rho_1 c_{con,k}(q;\rho_1;\rho_0)}.
\end{aligned}
\end{equation}
To make contact with the replica-symmetric formulation [see Eq.~(\ref{eq_ornstein_zernike_QA_RS})], we have defined in the above expressions $c_{con,k}(q;\rho;\rho_0)=\Gamma _{k1}^{(2)}(q;\rho;\rho_0)+\widehat \phi_k(q)$, $c _{mf,k}(q;\rho;\rho_0)=\Gamma _{k1}^{(1;1)}(q;\rho;\rho_0) $, and $c_{dis,k}( q;\rho_1, \rho_2;\rho_0)=\Gamma _{k2}^{(11)}(q;\rho_1, \rho_2;\rho_0)-\widetilde \phi_k(q)$. The disconnected direct correlation function $c_{dis,k}( q;\rho)$ of the replica-symmetric OZ equations is equal to $c_{dis,k}( q;\rho_1=\rho, \rho_2=\rho;\rho_0)$ in the above formalism; the correspondance is obvious for the two other direct correlation functions. 

\subsection{Asymptotic analysis}

As in the case of the naive HRT, the hierarchy simplifies when considered in the asymptotic regime $k,q\rightarrow 0$ near the critical point. We choose the IR regulators such that
\begin{equation}
\label{eq_hat_r_RSB}
\begin{aligned}
\widehat R_{k}( q)= Z_k k^2 \hat r(\hat q^2)
\end{aligned}
\end{equation}
and
\begin{equation}
\label{eq_tilde_r_RSB}
\begin{aligned}
\widetilde R_{k}( q)= K_k  \tilde r(\hat q^2)
\end{aligned}
\end{equation}
with $Z_k\sim k^{2-\eta}$, $K_k \sim k^{-(2\eta-\bar\eta)}$.

Similarly to what was done before, we introduce dimensionless quantities (after accounting for scaling dimensions that are appropriate near a zero-temperature fixed point): $\varphi_a \sim k^{-(d-4+\bar \eta)/2}(\rho_a - \rho_{crit})$, $ a_{k}(\varphi) \sim  k^{-(d-\theta)}[A_{f,k}(\rho;\rho_0) - A_{f,k;crit}(\rho;\rho_0)]$ with $\theta=2+\eta - \bar\eta$, etc. [We recall that $ \mathcal A_{f,k}(\rho;\rho_0)=\Gamma_{k1}(\rho;\rho_0)/V$.] We assume for simplicity in the following that $\bar \eta< 2 \eta$ so that all contributions involving matrix-fluid correlations are subdominant and can be neglected when $k\rightarrow 0$.

For a uniform density field, the first equation of the HRT hierarchy becomes
\begin{equation}
\label{eq_ERGE_a_k_asympt_RSB}
\begin{aligned}
&\partial_t a_{k}(\varphi)+ (d-2-\eta + \bar\eta) a_{k}(\varphi) - \frac{1}{2} (d-4+\bar \eta)\varphi \partial_{\varphi }a_{k}(\varphi)\\&= \frac{1}{2} \int_{\hat q}\bigg \{ \dot{\hat r}(\hat q^2)\frac{u_{dis,k}(\hat q^2;\varphi,\varphi)-\widetilde r(\hat q^2) }{[u_{con,k}(\hat q^2;\varphi)+ \hat r(\hat q^2)]^2}
+ \dot{\tilde r}(\hat q^2)\times \\& \frac{1}{[u_{con,k}(\hat q^2;\varphi)+\hat r(\hat q^2)]}\bigg \},
\end{aligned}
\end{equation}
where, to make a direct contact with the previous replica-symmetric treatment (see section~\ref{sec:naive}), we have defined 
\begin{equation}
\label{eq_con_u_RSB}
\begin{aligned}
u_{con,k}(\hat q^2;\varphi)\sim k^{-(2-\eta)} \Gamma_{k1}^{(2)}(q; \rho;\rho_0)
\end{aligned}
\end{equation}
and
\begin{equation}
\label{eq_dis_u_RSB}
\begin{aligned}
u_{dis,k}(\hat q^2;\varphi_1,\varphi_2)\sim k^{2\eta-\bar \eta}\Gamma_{k2}^{(11)}(q;\rho_1,\rho_2;\rho_0),
\end{aligned}
\end{equation}
with the dimensionless fields $\varphi_a$ and the densities $\rho_a$ related as above and $q=k\hat q$. (Note that we have chosen to not include the regulator $\tilde r$ in $u_{dis,k}$ to keep the dependence on $\tilde r$ explicit, despite the fact that as defined below Eq.~(\ref{eq_OZ_RSB_mf}), it is present in $c_{dis,k}$.) For compactness, we have also introduced the short-hand notations
\begin{equation}
\label{eq_partial_t_hat_r}
\dot{\hat r}(\hat q^2)=(2-\eta)\hat r(\hat q^2)- 2 \hat q^2 \hat r'(\hat q^2)
\end{equation}
and
\begin{equation}
\label{eq_partial_t_tilde_r}
\dot{\tilde r}(\hat q^2)=-(2\eta-\bar\eta )\tilde r(\hat q^2)- 2 \hat q^2 \tilde r'(\hat q^2).
\end{equation}
Note that all constants (that do not depend on $k$) can be incorporated in a trivial redefinition of the fields and functions.

Along the same lines, one obtains the flow of $v_k(\varphi_1,\varphi_2)\sim k^{-(d-2\theta)} [\Gamma_{k2}(\rho_1,\rho_2;\rho_0)/V]$ as
\begin{equation}
\label{eq_ERGE_v_k_asympt_RSB}
\begin{aligned}
&\partial_t v_{k}(\varphi_1,\varphi_2)+ (d-4-2\eta + 2\bar\eta) v_{k}(\varphi_1,\varphi_2) - \frac{1}{2} (d-4+\bar \eta)\\&\times (\varphi_1\partial_{\varphi_1 }+\varphi_2\partial_{\varphi_2})v_{k}(\varphi_1,\varphi_2)
\\&=-\frac{1}{2} \int_{\hat q} \bigg \{ \frac{\dot{\hat r}(\hat q^2)}{[u_{con,k}(\hat q^2;\varphi_1)+\hat r(\hat q^2)]}
\bigg (-\gamma_{k3}^{(110)}(\hat q^2;\varphi_1,\varphi_1,\varphi_2)
\\& + 2\, \gamma_{k2}^{(20)}(\hat q^2;\varphi_1,\varphi_2)\, \frac{[u_{dis,k}(\hat q^2;\varphi_1,\varphi_1)-\widetilde r(\hat q^2)]}{[u_{con,k}(\hat q^2;\varphi_1)+\hat r(\hat q^2)]^2} \\&
+ \frac{[u_{dis,k}(\hat q^2;\varphi_1,\varphi_2)-\widetilde r(\hat q^2)]^2}{[u_{con,k}(\hat q^2;\varphi_1)+\hat r(\hat q^2)][u_{con,k}(\hat q^2;\varphi_2)+\hat r(\hat q^2)]} \bigg ) +\\&
 \frac{\dot{\tilde r}(\hat q^2)}{[u_{con,k}(\hat q^2;\varphi_1)+\hat r(\hat q^2)]} 
\bigg (\frac{\gamma_{k2}^{(20)}(\hat q^2;\varphi_1,\varphi_2) }{[u_{con,k}(\hat q^2;\varphi_1)+\hat r(\hat q^2)]}
+\\& \frac{[u_{dis,k}(\hat q^2;\varphi_1,\varphi_2)-\widetilde r(\hat q^2)]}{[u_{con,k}(\hat q^2;\varphi_2)+\hat r(\hat q^2)]} \bigg )
+perm(12)\bigg \},
\end{aligned}
\end{equation}
where we have defined
\begin{equation}
\label{eq_gamma_2(20)dim}
\begin{aligned}
\gamma_{k2}^{(20)}(\hat q^2;\varphi_1,\varphi_2)\sim k^{2\eta-\bar \eta}\Gamma_{k2}^{(20)}(q;\rho_1,\rho_2;\rho_0)
\end{aligned}
\end{equation}
and
\begin{equation}
\label{eq_gamma_3(110)dim}
\begin{aligned}
\gamma_{k3}^{(110)}(\hat q^2;\varphi_1,\varphi_2,\varphi_3)\sim k^{2+ 3\eta-2 \bar \eta}\Gamma_{k3}^{(110)}(q;\rho_1,\rho_2, \rho_3;\rho_0),
\end{aligned}
\end{equation}
whereas $u_{dis,k}(\hat q=0;\varphi_1,\varphi_2)=\partial_{\varphi_1}\partial_{\varphi_2}v_k(\varphi_1,\varphi_2)$.

From the discussion in section~\ref{sec:superfield}, one knows that the underlying asymptotic supersymmetry implies Ward-Takahashi identities. In particular, for the the IR cutoff functions the identity in Eq.~(\ref{eq_ward_susy_momentum_2QA}) implies that $\widetilde R(q)= - K \partial_{q^2}\widehat R(q)$ with $K$ a constant. We therefore take the dimensionless cutoff functions such that
\begin{equation}
\label{eq_tilde_r_hat_r}
\begin{aligned}
\tilde r(\hat q^2)=-\partial_{\hat q^2}\hat r(\hat q^2)=-\hat r'(\hat q^2).
\end{aligned}
\end{equation}
The supersymmetry is then satisfied when $K_k=K Z_k$. The above choice nonetheless allows for a spontaneous breaking of the supersymmetry when $K_k\neq K Z_k$.

One can easily check that the above asymptotic HRT equations coincide with the nonperturbative RG flow equations for the RFIM\cite{tarjus04,tissier06,tissier11} obtained in the explicit replica-symmetry breaking framework. As shown in the latter studies, the flow equation involving the dimensionless second cumulant, Eq.~(\ref{eq_ERGE_v_k_asympt_RSB}), turns out to be a key element to resolve the conundrum of dimensional reduction.

\subsection{Supersymmetry-compatible OZ/LPA approximation}

We first reconsider the naive OZ/LPA approximation that we have used in the context of the replica-symmetric formalism (see section~\ref{subsec:naiveOZ/LPA}). Again, as there is no field renormalization and no disorder renormalization, $\eta = \bar \eta=0$, which implies that one cannot simply drop the matrix-fluid direct correlations in the asymptotic equations. However, we assume that no accidental cancellation takes place and that $c_{dis,k}(q)+\rho_0 S_{mm}(q)c_{mf,k}(q)^2\simeq K_k\rightarrow K_0$ in the small-$k$, small-$q$ limit. We then repeat the derivation of section~\ref{subsec:naiveOZ/LPA}, but we now consider two cutoff functions, $\widehat R_k$ and $\widetilde R_k$, which we can choose to be related so that their form does not explicitly break the underlying asymptotic supersymmetry. Accordingly, we use the Ward-Takahashi identity to require that $\widetilde R_k(q)= -K \partial_{q^2}\widehat R_k(q)$, with $K$ a constant to be determined later on. With the choice of regulator $\widehat R_k(q)=k^2 \hat r(q^2/k^2)$, one therefore has $\widetilde R_k(q)= -K \hat r'(q^2/k^2)$.

The asymptotic form of the HRT equation for the Helmholtz free-energy density $a_k(\varphi)$ now reads:
\begin{equation}
\label{eq_ERGE_a_k_asympt_LPA}
\begin{aligned}
&\partial_t a_{k}(\varphi)+ (d-2) a_{k}(\varphi) - \frac{1}{2} (d-4)\varphi \partial_{\varphi }a_{k}(\varphi)=\\& \frac{K_0}{2}\int_{\hat q}  \frac{2[\hat r(\hat q^2)- \hat q^2 \hat r'(\hat q^2)]}{[u_{con,k}(\hat q^2;\varphi)+ \hat r(\hat q^2)]^2}+\frac{K}{2}\times \\& \int_{\hat q} \bigg\{ \frac{2\hat q^2 \hat r''(\hat q^2)}{u_{con,k}(\hat q^2;\varphi)+ r(\hat q^2)}+\frac{2[\hat r(\hat q^2)- \hat q^2 \hat r'(\hat q^2)]\hat{r}'(\hat q^2)}{[u_{con,k}(\hat q^2;\varphi)+ \hat r(\hat q^2)]^2}\bigg\}
\end{aligned}
\end{equation}
with $u_{con,k}(\hat q^2)= \hat q^2 + a''_k(\varphi)$.

Let us introduce the variable $y=\hat q^2$ and define as before $v_d=[2^{d+1}\pi^{d/2}\Gamma(d/2)]^{-1}$ [it is equal to $1/4$th of the area of the $d$-dimensional sphere of radius unity divided by $(2\pi)^d$]. Then, after using the fact that $\partial_y [u_{con,k}(y;\varphi)+ \hat r(y)]=1+\hat r'(y)$, one finds
\begin{equation}
\label{eq_ERGE_a_k_asympt_LPA_proof1}
\begin{aligned}
&\partial_t a_{k}(\varphi)+ (d-2) a_{k}(\varphi) - \frac{1}{2} (d-4)\varphi \partial_{\varphi }a_{k}(\varphi)=\\&  v_d(K_0-K)\int dy \, y^{\frac{d}{2}-1} \frac{2[\hat r(y)- y \hat r'(y)]}{[u_{con,k}(y;\varphi)+ \hat r(y)]^2}\\& - v_d\, K  \int dy\, y^{\frac{d}{2}-1}\partial_y\left\{ \frac{2[\hat r(y)- y \hat r'(y)]}{[u_{con,k}(y;\varphi)+ \hat r(y)]}\right\}.
\end{aligned}
\end{equation}
After integration by parts, the second term of the right-hand side can be reexpressed as
\begin{equation}
\label{eq_ERGE_a_k_asympt_LPA_proof2}
\begin{aligned}
&- v_d\, K  \int dy\, y^{\frac{d}{2}-1}\partial_y\left\{ \frac{2[\hat r(y)- y \hat r'(y)]}{[u_{con,k}(y;\varphi)+ \hat r(y)]}\right\}\\&=K\, v_d \left(\frac{d-2}{2}\right)\int dy\, y^{\frac{d-2}{2}-1}\frac{2[\hat r(y)- y \hat r'(y)]}{[u_{con,k}(y;\varphi)+ \hat r(y)]}\\&
=\left (\frac{K}{4\pi}\right )\frac{1}{2}\int_{\hat q}^{(d-2)} \frac{2[\hat r(\hat q^2)- \hat q^2 \hat r'(\hat q^2)]}{u_{con,k}(\hat q^2;\varphi)+ \hat r(\hat q^2)},
\end{aligned}
\end{equation}
where the integral over the wavevector $\hat q$ is now in a $(d-2)$-dimensional space. In deriving the above equalities we have used the fact that $v_d (d-2)/2=v_{d-2}/(4\pi)$.

If we now choose $K=K_0$,  the HRT equation for $a_k(\varphi)$ reduces to
\begin{equation}
\label{eq_ERGE_a_k_asympt_LPA_proof3}
\begin{aligned}
&\partial_t a_{k}(\varphi)+ (d-2) a_{k}(\varphi) - \frac{1}{2} (d-4)\varphi \partial_{\varphi }a_{k}(\varphi)=\\& \left (\frac{K_0}{4\pi}\right )\frac{1}{2}\int_{\hat q}^{(d-2)} \frac{2[\hat r(\hat q^2)- \hat q^2 \hat r'(\hat q^2)]}{\hat q^2+a''_k(\varphi)+ \hat r(\hat q^2)},
\end{aligned}
\end{equation}
which, up to the trivial constant factor $K_0/(4\pi)$, is exactly the HRT equation for a pure fluid in the OZ/LPA approximation in dimension $d-2$ [compare with Eq.~(\ref{eq_ERGE_A_k_asympt_naiveOZ})]. Within this approximation, the exponents describing the gas-liquid critical point of a fluid in a disordered porous material are therefore identical to those of the pure fluid in two dimensions less. For instance, for the sharp cutoff in $d=5$, this provides the exponents $\nu\simeq 0.689$, $\gamma \simeq1.378$, $\delta=5$, and $\omega \simeq 0.581$ as obtained for the bulk fluid in $d=3$ in the same approximation and with the same sharp cutoff\cite{parola-reatto85,HRTreview}. This proves that dimensional reduction is obeyed within this OZ/LPA approximation, provided that one properly chooses the infrared regulators. The results from the naive OZ/LPA approximation were thus plainly wrong.

\subsection{Improved truncations}

The OZ/LPA approximation, even when designed not to break the supersymmetry explicitly, is not satisfactory. First, it does not allow one to investigate a \textit{spontaneous} breaking of supersymmetry and the associated breakdown of dimensional reduction at the critical point, and second, it prevents one from studying dimensions less than $4$. Improved truncations of the HRT hierarchy are therefore required which at least include the second cumulant of the renormalized disorder for arguments that are generically different (so that a nonanalyticity can freely emerge along the RG flow, see above) and provide a better description of the long-distance behavior of the correlations.

From the nonperturbative RG work on the RFIM\cite{tarjus04,tissier06,tissier11}, we have learned that a truncation must include the first cumulant $\Gamma_{k1}[\rho;\rho_0]$ and the second cumulant $\Gamma_{k2}[\rho_1,\rho_2; \rho_0]$. In the HRT framework, which is operationally based on coupled flow equations for 1PI vertex functions evaluated for uniform density fields, this implies approximation schemes to the two-point (or higher-order) direct correlation functions obtained from $\Gamma_{k1}$ and $\Gamma_{k2}$. Interestingly, the formalism with explicit breaking of the replica symmetry that we now use provides us with additional ``compressibility sum rules''. At the level of the direct pair correlation functions, in addition to Eqs.~(\ref{eq_compress_sum_rule_con}) and (\ref{eq_compress_sum_rule_mf}) relating the connected fluid-fluid pair correlations and the matrix-fluid ones to the first cumulant, \textit{i.e.}, the Helmoltz free-energy density of the fluid $\mathcal A_{f,k}(\rho;\rho_0)=\Gamma_{k1}(\rho;\rho_0)/V$, there is a new sum rule for the disconnected fluid-fluid direct pair correlation function,
\begin{equation}
\label{eq_compress_sum_rule_dis}
\begin{aligned}
&c_{dis,k}(q=0;\rho_1,\rho_2;\rho_0)-\widetilde \phi_k(q=0)\\&=\Gamma_{k2}^{(11)}(\rho_1,\rho_2; \rho_0)=\frac{\partial^2 V_k(\rho_1,\rho_2;\rho_0)}{\partial\rho_1 \partial \rho_2} ,
\end{aligned}
\end{equation}
where $V_k(\rho_1,\rho_2;\rho_0)=\Gamma_{k2}(\rho_1,\rho_2; \rho_0)/V$ is the second cumulant for uniform fields (divided by the sample volume $V$ whose notation should not be confused with $V_k$).

A minimal truncation of the HRT hierarchy should thus include $\mathcal A_{f,k}(\rho_1;\rho_0)$, $V_k(\rho_1,\rho_2;\rho_0)$, $c_{con,k}(q;\rho_1;\rho_0)$, $c_{mf,k}(q;\rho_1;\rho_0)$, and $c_{dis,k}(q;\rho_1,\rho_2;\rho_0)$, with sum rules relating the three direct correlation functions at $q=0$ to the two potentials. The remaining challenge is to go beyond the OZ/LPA treatment of the spatial or $q$ dependence of the correlation functions. In the field-theoretical framework, there are two ways to achieve this and, in consequence, to predict nonzero anomalous dimension(s). The standard procedure is to consider next orders of the ``derivative expansion'', which is an expansion of the effective average action in gradients of the field\cite{berges02}. A recently implemented alternative consists of truncating the hierarchy of equation for the 1PI vertices by approximating the momentum dependence of the $3$- and $4$-point vertices so that the equation for the $2$-point vertex can be closed through a systematic use of compressiblity-like sum rules\cite{BMW}. (This is akin to a procedure formulated earlier by Parola and Reatto\cite{HRTreview} for fluids but never implemented in this context.) In the context of the RFIM, it has however proven difficult to use this last approximation as it explicitly breaks the underlying supersymmetry\cite{tissier11}.

It appears that a possible candidate for a practical implementation of the HRT for describing fluids in disordered porous materials from short- to long-distance physics would be a combination of the derivative expansion for small wavevectors with an OZ-like approximation carrying all microscopic details for high wavevectors. Dropping for simplicity the explicit $\rho_0$ dependence in the direct correlation functions, a typical approximation would then read
\begin{equation}
\label{eq_closure_conn_HRT_RSB}
\begin{aligned}
&\frac{1}{\rho}- c_{con,k}(q;\rho)-\widehat \phi_k(q)=\\& \frac{\partial^2\mathcal A_{f,k}(\rho;\rho_0)}{\partial \rho^2}+ Z_k(\rho;\rho_0)q^2 \Theta(q-q^\dag) -[1-\Theta(q-q^\dag)]\times \\& \left[c_{con,R}(q)-c_{con,R}(q=0) +\phi(q)-\phi(q=0) \right ],
\end{aligned}
\end{equation}
where $\Theta(q-q^\dag)$ is a smoothed Heaviside step function that goes continuously from $1$ when $q<q^\dag$  to $0$ when $q>q^\dag$ with $q^\dag$ a crossover wavevector larger than the IR cutoff $k$ and smaller than the typical interatomic distance in the liquid, $\sigma$ [one could also use an approximate form in the spirit of Eq.~(\ref{eq_RPAbis})], and
\begin{equation}
\label{eq_closure_mf_HRT_RSB}
\begin{aligned}
c_{mf,k}(q;\rho)=-&\frac{\partial^2\mathcal A_{f,k}(\rho;\rho_0)}{\partial \rho \partial \rho_0}\\&+ [c_{mf,R}(q;\rho) -c_{mf,R}(q=0;\rho)],
\end{aligned}
\end{equation}
\begin{equation}
\label{eq_closure_dis_HRT_RSB}
\begin{aligned}
c_{dis,k}(q;\rho_1,\rho_2)-\widetilde \phi_k(q)=&\frac{\partial^2 V_k(\rho_1,\rho_2)}{\partial\rho_1 \partial \rho_2}+ [c_{dis,R}(q;\rho_1,\rho_2)\\& -c_{dis,R}(q=0;\rho_1,\rho_2)].
\end{aligned}
\end{equation}
Again, one can improve this ansatz by implementing a core condition for the total fluid-fluid and the matrix-fluid pair correlations. In addition, we set to zero the third and the higher-order cumulants, so that the HRT hierarchy is closed.

In the asymptotic regime, when $k,q \rightarrow 0$ (and if $\bar \eta<2\eta$ so that one can drop the matrix-fluid contribution), one has $u_{con,k}(\hat q)=a_k''(\varphi)+z_k(\varphi)\hat q^2$ and $u_{dis,k}(\hat q)=\partial_{\varphi_1}\partial_{\varphi_2}v_k(\varphi_1,\varphi_2)$. It is easy to show\cite{tissier11} that the supersymmetry is satisfied when $\partial_{\varphi_1}\partial_{\varphi_2}v_k(\varphi_1,\varphi_2)\vert_{\varphi_1=\varphi_2=\varphi}=z_k(\varphi)$ and (spontaneously) broken otherwise. This truncation has been shown to provide a very good description of the critical behavior of the RFIM in $d=3$ while resolving the dimensional-reduction puzzle\cite{tissier11}.

With the above closure, one has to solve, on top of the OZ equations, three coupled nonlinear partial differential equations describing the flows of $\mathcal A_{f,k}(\rho;\rho_0)$, $V_k(\rho_1,\rho_2;\rho_0)$ and $Z_k(\rho;\rho_0)$, the latter being obtained from $\partial_{q^2}\Gamma_{k1}^{(2)}(q^2;\rho;\rho_0)\vert_{q=0}$. The solution would allow a complete description, at all length scales, of a fluid in a disordered porous medium in $d=3$. However, this represents a very arduous numerical task\cite{footnoteVI}.

\section{conclusion}
\label{sec:conclusion}

In this article we have considered the equilibrium behavior of fluids adsorbed in disordered mesoporous materials, with a special attention to the gas-liquid critical point. Because it allows one to keep track of the details of the physics at a microscopic level while dealing with fluctuations on arbitrarily large length scales, the HRT provides an efficient framework. A nontrivial generalization of the HRT of bulk fluids and mixtures is however required. Through a description of the fluid/matrix system as a quenched-annealed mixture, we have combined liquid-state statistical mechanics and formalism borrowed from the theory of systems with quenched disorder. 

A straightforward implementation of the HRT to the replica description of the quenched-annealed mixture, which in particular assumes replica symmetry, has been shown to lead to results concerning the critical behavior that are inconsistent and unsatisfactory. At the same time, it has provided strong indication that the critical behavior of the quenched-annealed mixture and that of the RFIM are in the same universality class. The same difficulties encountered in the RG treatment of the latter are then to be expected in the HRT of the former. We have then built on the recent nonperturbative RG approach of the RFIM developed by two of us for solving pending puzzles in the long-distance behavior of the model\cite{tarjus04,tissier06,tissier11} to propose a more sophisticated HRT for the quenched-annealed mixture. As discussed and illustrated in the paper, this formalism opens the way to cure the inconsistencies of the naive treatment. We have also devised approximations to the HRT hierarchy that are expected to lead to an accurate description of the behavior of fluids in disordered porous materials. A full-blown resolution of the resulting set of coupled nonlinear differential equations is however a formidable task that we defer to future work.

\appendix

\section{Higher orders of the naive HRT hierarchy in the asymptotic regime}
\label{app:asymptotic_naive}

We have derived in section~\ref{sub:asymptotic} the asymptotic equation for the Helmholtz free-energy density and we have introduced scaling dimensions appropriate for studying a zero-temperature fixed point. The scaling of the $2$-point correlation function functions is given in Eqs.~(\ref{eq_vertex2conn_asymptotic_naive}) and (\ref{eq_vertex2dis_asymptotic_naive}). Here, as often in the main text, we assume for simplicity that $\bar \eta < 2 \eta$, which implies that the contribution from the matrix-fluid  correlations always lead to subdominant terms in the asymptotic regime.

Some care is needed to derive the proper scaling of the higher-order direct correlation functions as there are \textit{a priori} an increasing number of distinct functions when $n\rightarrow 0$ (whereas only the connected and disconnected components appear at the pair level). For instance, the 3-body direct correlation functions should scale as
\begin{equation}
\label{eq_vertex31_asymptotic_naive}
\begin{aligned}
\lim_{n\rightarrow 0}\sum_{a,b=1}^n&\Gamma_{k;1ab}^{rep(3)}(q_1,q_2,q_3)\\& \simeq k^{d-\theta-\frac{3}{2}(d-4+\bar \eta)}\, u_{1,k}(\hat q_1,\hat q_2,\hat q_3)
\end{aligned}
\end{equation}
\begin{equation}\label{eq_vertex32_asymptotic_naive}
\begin{aligned}
\lim_{n\rightarrow 0}\sum_{a=1}^n&\Gamma_{k;12a}^{rep(3)}(q_1,q_2,q_3)\\& \simeq - k^{d-2\theta-\frac{3}{2}(d-4+\bar \eta)}\, u_{12,k}(\hat q_1,\hat q_2,\hat q_3)
\end{aligned}
\end{equation}
\begin{equation}\label{eq_vertex33_asymptotic_naive}
\begin{aligned}
\lim_{n\rightarrow 0}&\Gamma_{k;123}^{rep(3)}(q_1,q_2,q_3)\\& \simeq k^{d-3\theta-\frac{3}{2}(d-4+\bar \eta)}\, u_{123,k}(\hat q_1,\hat q_2,\hat q_3)
\end{aligned}
\end{equation}
Similarly,
\begin{equation}
\label{eq_vertex41_asymptotic_naive}
\begin{aligned}
\lim_{n\rightarrow 0}\sum_{a,b,c=1}^n&\Gamma_{k;1abc}^{rep(4)}(q_1,q_2,q_3,q_4)\\& \simeq k^{d-\theta-2(d-4+\bar \eta)}\, w_{1,k}(\hat q_1,\hat q_2,\hat q_3,\hat q_4)
\end{aligned}
\end{equation}
\begin{equation}
\label{eq_vertex42_asymptotic_naive}
\begin{aligned}
\lim_{n\rightarrow 0}\sum_{a,b=1}^n&\Gamma_{k;12ab}^{rep(4)}(q_1,q_2,q_3,q_4)\\& \simeq - k^{d-2\theta-2(d-4+\bar \eta)}\, w_{12,k}(\hat q_1,\hat q_2,\hat q_3,\hat q_4),
\end{aligned}
\end{equation}
etc., with $\theta=2+\eta -\bar \eta$. The expressions of the various functions appearing in the above equations can actually be rationalized by making use of the cumulants of the renormalized disorder that can be introduced when explicitly breaking the replica symmetry (see section~\ref{sub:explicitRSB} and the following appendix). This connection also justifies the choice of sign used in the above equations. 

We illustrate the higher-order equations of the naive HRT hierarchy for case of the connected direct pair correlation function ($2$-point 1PI vertex). It reads
\begin{equation}
\label{eq_ERGE_u_con_asympt_naive}
\begin{aligned}
&\partial_t u_{con,k}(\hat q^2;\varphi)+ (2-\eta) u_{con,k}(\hat q^2;\varphi)- \frac{1}{2} (d-4+\bar \eta)\times \\&\big (\varphi  \partial_{\varphi }+\sum_{\mu=1}^d \hat q^{\mu}\partial_{\hat q^{\mu}}\big )u_{con,k}(\hat q^2;\varphi) =
\\&- \frac{1}{2}\widetilde\partial_t  \int_{\hat q'} \bigg \{ -\frac{w_{1,k}(\hat q',-\hat q',\hat q,-\hat q)\, u_{dis,k}(\hat q'^2)}{[u_{con,k}(\hat q'^2)+ r(\hat q'^2)]^2}\;+\\&
2\, \frac{[u_{1,k}(\hat q,\hat q',-(\hat q+\hat q'))]^2 \,u_{dis,k}(\hat q'^2)}{[u_{con,k}(\hat q'^2)+ r(\hat q'^2)]^2[u_{con,k}(\vert \hat q+\hat q'\vert^2)+ r(\vert \hat q+\hat q'\vert^2)]}\\&
-2\, \frac{u_{12,k}(\hat q',-(\hat q +\hat q'),\hat q)\, u_{1,k}(\hat q',-(\hat q +\hat q'),\hat q) }{[u_{con,k}(\hat q'^2)+ r(\hat q'^2)][u_{con,k}(\vert \hat q+\hat q'\vert^2)+ r(\vert \hat q+\hat q'\vert^2)]}\\&
 + \frac{w_{12,k}(\hat q',-\hat q',\hat q,-\hat q) }{[u_{con,k}(\hat q'^2)+ r(\hat q'^2)]}\bigg\},
\end{aligned}
\end{equation}
where the dependence on the dimensionless field $\varphi$ is not explicitly shown in the right-hand side and we recall that the operator $\widetilde \partial_t$ only acts on the regulator, with the formal definition $\widetilde \partial_t r(\hat q^2)\equiv \dot r(\hat q^2)\equiv (2-\eta)r(\hat q^2)-2\hat q^2 r'(\hat q^2)$. Similar equations can also be derived for the flow of $u_{dis,k}(\hat q^2;\varphi)$ and all higher-order dimensionless 1PI vertices but it is not worth displaying them here.

The existence of ``compressibility sum rules'' and their generalizations, which merely express the fact that the effective action is the generating functional of the 1PI vertices, allows one to identify the above equation for $u_{con,k}(\hat q^2;\varphi)$ when $\hat q^2=0$ with the second derivative with respect to the field $\varphi$ of the flow equation for $a_k(\varphi)$,  Eq.~(\ref{eq_ERGE_A_k_asympt_naive}). With the help of the operator $\widetilde \partial_t$, Eq.~(\ref{eq_ERGE_A_k_asympt_naive}) can be reexpressed as 
\begin{equation}
\label{eq_ERGE_A_k_asympt_naive_tilde}
\begin{aligned}
&\partial_t a_{k}(\varphi)+ (d-2-\eta + \bar\eta) a_{k}(\varphi) - \frac{1}{2} (d-4+\bar \eta)\varphi \partial_{\varphi }a_{k}(\varphi)\\&= -\frac{1}{2} \widetilde \partial_t \int_{\hat q} \frac{u_{dis,k}(\hat q^2) }{[u_{con,k}(\hat q^2)+ r(\hat q^2)]}.
\end{aligned}
\end{equation}
Then, by deriving twice this equation with respect to $\varphi$, one obtains
\begin{equation}
\label{eq_ERGE_A_ksecond_asympt_naive_tilde}
\begin{aligned}
&\partial_t a''_{k}(\varphi)+ (2-\eta) a''_{k}(\varphi) - \frac{1}{2} (d-4+\bar \eta)\varphi \partial_{\varphi }a''_{k}(\varphi)= \\& 
-\frac{1}{2} \widetilde \partial_t \int_{\hat q} \bigg \{-\frac{u_{dis,k}(\hat q^2) \partial_{\varphi}^2 u_{con,k}(\hat q^2) }{[u_{con,k}(\hat q^2)+ r(\hat q^2)]^2}+\frac{\partial_{\varphi}^2 u_{dis,k}(\hat q^2) }{[u_{con,k}(\hat q^2)+ r(\hat q^2)]}
\\& + 2 \frac{u_{dis,k}(\hat q^2) [\partial_{\varphi} u_{con,k}(\hat q^2)]^2 }{[u_{con,k}(\hat q^2)+ r(\hat q^2)]^3}
-2 \frac{\partial_{\varphi}  u_{dis,k}(\hat q^2) \partial_{\varphi} u_{con,k}(\hat q^2) }{[u_{con,k}(\hat q^2)+ r(\hat q^2)]^2}
\bigg \}
\end{aligned}
\end{equation}
where again the $\varphi$ dependence is omitted in the right-hand side. The compressibility sum rule for the connected fluid-fluid pair correlations states that $u_{con,k}(\hat q^2=0;\varphi)=\partial_{\varphi}^2 a_k(\varphi)=a''_k(\varphi)$. Then, by setting $\hat q^2=0$ in Eq.~(\ref{eq_ERGE_u_con_asympt_naive}) and using the generalized  sum rules,  $u_{1,k}(0,\hat q,-\hat q)=\partial_{\varphi} u_{con,k}(\hat q^2)$, $u_{12,k}(\hat q,-\hat q,0)=\partial_{\varphi}  u_{dis,k}(\hat q^2) $, $w_{1,k}(\hat q,-\hat q, 0,0)=\partial_{\varphi}^2 u_{con,k}(\hat q^2)$, and $w_{12,k}(\hat q,-\hat q,0,0)=\partial_{\varphi}^2 u_{dis,k}(\hat q^2)$, one recovers Eq.~(\ref{eq_ERGE_A_ksecond_asympt_naive_tilde}).

\section{Algebraic manipulations for the expansions in number of free replica sums}
\label{app:algebraic_replica_sums}

We consider the replicated version of the quenched-annealed mixture in which the chemical potentials are different for each fluid replica so that the density fields $\{\rho_a\}$ are also different. The effective action (Helmholtz free-energy functional) for the $(n+1)$-component mixture can be expanded in increasing number of free replica sums as indicated in Eq.~(\ref{eq_Gamma_cumulants}). The developments below generalize to the $(n+1)$-component system the results for the RFIM\cite{tarjus04}; in the latter the average over the random field has been performed so that one is left with an $n$-component mixture with, however, interactions between the replicas.

Consider a generic (symmetric) matrix $A_{\alpha \beta}[\left\lbrace  \rho_f\right\rbrace;\rho_0 ]$, where we have denoted $\left\lbrace  \rho_f\right\rbrace$ the $n$ replica density fields to avoid confusion in the indices (we recall that Greek indices denote any component, including the matrix, whereas Roman indices denote fluid replicas only). Its $n\times n$ submatrix  associated with the fluid replicas can be decomposed as
\begin{equation}
\label{eq_genericdecompos}
A_{ab}[\left\lbrace  \rho_f\right\rbrace; \rho_0  ]=\widehat{A}_{a}[\left\lbrace  \rho_f\right\rbrace; \rho_0  ] \delta_{ab}+ \widetilde{A}_{ab}[\left\lbrace  \rho_f\right\rbrace ; \rho_0 ].
\end{equation}
In the above expression,  it is understood that the second term $\widetilde{A}_{ab}$ no longer contains any Kronecker symbol.
Each component of the matrix $\vect A$ can now be expanded in increasing number of free replica sums,
\begin{equation}
\label{eq_A00}
A_{00}[\left\lbrace  \rho_f\right\rbrace; \rho_0 ]=A_{00}^{[0]}[ \rho_0 ]+\sum_{c=1} ^{n}A_{00}^{[1]}[ \rho_c; \rho_0  ]+\cdots
\end{equation}
\begin{equation}
\label{eq_A0a}
A_{0a}[\left\lbrace  \rho_f\right\rbrace; \rho_0   ]=A_{01}^{[0]}[ \rho_a; \rho_0   ]+\sum_{c=1} ^{n}A_{01}^{[1]}[ \rho_a \vert  \rho_c; \rho_0    ]+\cdots
\end{equation}
\begin{equation}
\label{eq_widehatA}
\widehat{A}_{a}[\left\lbrace  \rho_f\right\rbrace; \rho_0   ]=\widehat{A}^{[0]}[ \rho_a; \rho_0   ]+\sum_{c=1} ^{n}\widehat{A}^{[1]}[ \rho_a \vert  \rho_c; \rho_0    ]+\cdots
\end{equation}
\begin{equation}
\label{eq_widetildeA}
\widetilde{A}_{ab}[\left\lbrace  \rho_f\right\rbrace; \rho_0   ]=\widetilde{A}^{[0]}[ \rho_a,  \rho_b; \rho_0  ]+\sum_{c=1} ^{n}\widetilde{A}^{[1]}[ \rho_a,  \rho_b \vert   \rho_c; \rho_0  ]+\cdots,
\end{equation}
where the superscripts in square brackets denote the order in the expansion (and should not be confused with superscripts in parentheses indicating partial derivatives).

As an illustration, the expansion of the matrix $\Gamma_{k}^{(2)}$ of $2$-point 1PI vertices (direct correlation functions) reads in terms of the expansion of effective average action itself (for ease of notation we drop the subscript $k$ in the expressions appearing in the right-hand sides):
\begin{equation}
\label{eq_Gamma00}
\Gamma_{k;00}^{(2)}[\left\lbrace  \rho_f\right\rbrace; \rho_0 ]=\Gamma_{0}^{(2)}[ \rho_0 ]+\sum_{c=1} ^{n}\Gamma_{1}^{(0;1)}[ \rho_c; \rho_0  ]+\cdots
\end{equation}
\begin{equation}
\label{eq_Gamma0a}
\Gamma_{k;0a}^{(2)}[\left\lbrace  \rho_f\right\rbrace; \rho_0   ]=\Gamma_{1}^{(1;1)}[ \rho_a; \rho_0   ]-\sum_{c=1} ^{n}\Gamma_{2}^{(10;1)}[ \rho_a, \rho_c; \rho_0    ]+\cdots
\end{equation}
\begin{equation}
\label{eq_widehatGamma}
\widehat{ \Gamma}_{k;a}^{(2)}[\left\lbrace  \rho_f\right\rbrace; \rho_0  ]= \Gamma_{1}^{(2)}[ \rho_a; \rho_0  ]-\sum_{c=1} ^{n} \Gamma_{2}^{(20)}[ \rho_a,  \rho_c; \rho_0   ]+\cdots
\end{equation}
\begin{equation}
\label{eq_widetildeGamma}
\begin{split}
\widetilde{ \Gamma}_{k;ab}^{(2)}[\left\lbrace  \rho_f\right\rbrace; \rho_0  ]=-  \Gamma_{2}^{(11)}& [ \rho_a,  \rho_b; \rho_0 ]\\& +\sum_{c=1} ^{n} \Gamma_{3}^{(110)}[ \rho_a,  \rho_b,  \rho_c; \rho_0 ]+\cdots,
\end{split}
\end{equation}
where the permutational symmetry of the arguments of the $\Gamma_{kp}$'s has been used and the superscripts indicate the number of functional derivatives with respect to the density fields appearing as arguments. For simplicity, we have not used the superscript $0$ when there is no derivative with respect to $\rho_0$ and we have not indicated the dependence on spatial coordinates or momenta.

The above expressions can be used to reinterpret the various direct correlation functions (1PI vertices) appearing in the replica-symmetric formalism. We focus here on the fluid-fluid correlations, extension to the matrix-fluid ones being straightforward. One readily obtains that $c_{con,k}(q,\rho)$ is the same in the two formalisms and is related to a $2$-point 1PI vertex obtained from the first cumulant, $\Gamma_{k1}^{(2)}(q;\rho)$; $c_{dis,k}(q;\rho)$ in the replica-symmetric formalism is equal to $c_{dis,k}(q;\rho_1=\rho,\rho_2=\rho)$, which is related to a $2$-point 1PI vertex obtained from the second cumulant, $\Gamma_{k2}^{(11)}(q;\rho,\rho)$, in the formalism with explicit replica symmetry breaking (for $c_{con,k}$ and $c_{dis,k}$ one also has to include IR regulators).

Similarly, the higher-order vertices appearing in the previous appendix can be expressed as 
\begin{equation}
\label{eq_vertex31_RS_RSB}
\begin{aligned}
\lim_{n\rightarrow 0} \sum_{a,b=1}^n \Gamma_{k;1ab}^{rep(3)}(q_1,q_2,q_3;\rho)=\Gamma_{k1}^{(3)}(q_1,q_2,q_3;\rho),
\end{aligned}
\end{equation}
\begin{equation}
\label{eq_vertex32_RS_RSB}
\begin{aligned}
&\lim_{n\rightarrow 0} \sum_{a=1}^n\Gamma_{k;12a}^{rep(3)}(q_1,q_2,q_3;\rho)\\&=-[\Gamma_{k2}^{(21)}(q_1,q_2,q_3;\rho,\rho)+\Gamma_{k2}^{(12)}(q_1,q_2,q_3;\rho,\rho)] ,
\end{aligned}
\end{equation}
\begin{equation}
\label{eq_vertex33_RS_RSB}
\begin{aligned}
\lim_{n\rightarrow 0}\Gamma_{k;123}^{rep(3)}(q_1,q_2,q_3;\rho)=\Gamma_{k3}^{(111)}(q_1,q_2,q_3;\rho,\rho,\rho),
\end{aligned}
\end{equation}
etc.

Algebraic manipulations on matrices such as those defined above can be performed by a term-by-term identification of the orders of the expansions. For instance, the inverse $\vect B=\vect A^{-1}$ of the matrix $\vect A$ can also be put in the form of Eq.~(\ref{eq_genericdecompos}) and its components can be expanded in number of free replica sums. The term-by-term identification of the condition $\vect A \vect B= \vect 1$ leads to a \textit{unique} expression of the various orders,  $B_{00}^{[p]}$, $B_{01}^{[p]}$, $\widehat{B}^{[p]}$, and $\widetilde{B}^{[p]}$, of the expansion of $\vect B$  in terms of the $A_{00}^{[q]}$'s, $A_{01}^{[q]}$'s, $\widehat{A}^{[q]}$'s and $\widetilde{A}^{[q]}$'s with $q\leq p$. The algebra becomes rapidly tedious, and we only illustrate here the results for the zeroth-order terms:
\begin{equation}
\label{eq_B00_0}
B_{00}^{[0]}[\rho_0 ]=A_{00}^{[0]}[ \rho_0 ]^{-1},
\end{equation}
\begin{equation}
\label{eq_B0a_0}
B_{0a}^{[0]}[ \rho_1; \rho_0   ]=-B_{00}^{[0]}[\rho_0 ]A_{01}^{[0]}[ \rho_1; \rho_0   ]\widehat{B}^{[0]}[  \rho_1; \rho_0],
\end{equation}
\begin{equation}
\label{eq_widehatB_0}
\widehat{B}^{[0]}[  \rho_1; \rho_0] = \widehat{A}^{[0]}[  \rho_1; \rho_0]^{-1},
\end{equation}
\begin{equation}
\begin{aligned}
\label{eq_widetildeB_0}
\widetilde{B}^{[0]}[  \rho_1, \rho_2; \rho_0]& = - \widehat{B}^{[0]}[  \rho_1; \rho_0]\bigg (\widetilde{A}^{[0]}[  \rho_1, \rho_2; \rho_0]-\\&A_{10}^{[0]}[\rho_1; \rho_0]B_{00}^{[0]}[\rho_0]A_{10}^{[0]}[\rho_2; \rho_0] \bigg )\widehat{B}^{[0]}[  \rho_2; \rho_0].
\end{aligned}
\end{equation}
Note that by combining Eq.~(\ref{eq_Gamma00}) and Eq.~(\ref{eq_B00_0}), one can immediately see that for a uniform density field, $\Gamma_{k,00}^{(2)[0]}(q;\rho_0)$ is simply $1/[\rho_0 S_{mm}(q;\rho_0)]$ and is independent of the presence of the fluid.

\acknowledgments

It is a pleasure to dedicate this article to professor Luciano Reatto whose work has been strongly influential to our study of critical behavior in quenched disordered environments.


\begin{thebibliography}{15}
\bibitem{footnote-dimred}
This comes from the fact that the HRT equation in the OZ/LPA approximation for the bulk fluid with a sharp cutoff reads

$\partial_t a_{*}(\varphi)=-d a_{*}(\varphi) +\frac{1}{2} (d-2)\varphi a'_{*}(\varphi) - 
\log\left (\frac{1+a''_{*}(\varphi)}{1+a''_{*}(0)}\right )$

and is not the same as that for the quenched-annealed mixture in dimension $d+2$, Eq.~(\ref{eq_ERGE_a_k_naiveOZ_sharp}).
\bibitem{footnote-matrix}
One may indeed think of the quenched matrix configurations as obtained from an equilibrium system with a possibly complicated  interaction energy function; contrary to what is sometimes assumed for simplicity, this energy function is not pairwise additive in general and involves irreducible many-body terms, whose description is irrelevant here. Then, the corresponding direct correlation functions as related to the standard (total or Green's) correlation functions through a Legendre transform. $\Gamma_0$, which is a functional of a solid-density field $\rho_0(x)=<\hat \rho_0(x)>$, is only introduced as a convenient trick to keep track of the structural information about the matrix the matrix. In practice, only the static structure factor is needed.
\bibitem{footnoteVI}
As a warm-up, it could be worthwhile to solve the full HRT equations (not only their asymptotic form) with an OZ-like closure (RPA or ORPA) for the quenched-annealed mixture in $d=5$ with the proper set of IR regulators. As we have shown, this approximation predicts that the liquid-gas critical behavior is the same as that of the bulk fluid in $d=3$. While such a dimensional-reduction property is not expected to be exact for $d<d_{DR}\simeq 5.1$\cite{tissier11}, the error should be small in $d=5$. On the other hand, this closure, which incorporates a microscopic description of confinement, randomness and wettability, would allow one to study crossover phenomena determined by the competition between different fixed points (RFIM, bulk, site-diluted Ising model, percolation) and confirm if the liquid-gas critical behavior in the presence of a disordered matrix is indeed in the universality class of the RFIM.
\end{thebibliography}
\end{document}